\documentstyle[12pt]{article}
\def\beq{\begin{equation}}
\def\eeq{\end{equation}}
\def\bea{\begin{eqnarray}}
\def\eea{\end{eqnarray}}
\def\dslash{\partial{\!\!\!/}}
\def\adot{\dot{a}}
\def\addot{\ddot{a}}

\newcommand\mpl{M_{\rm P}}

\catcode`\@=11
\def\marginnote#1{}
\def\ifmath#1{\relax\ifmmode #1\else $#1$\fi}

\def\dslash{\partial\llap{/}}
\def\hgam{{\hat \gamma}}
\def\adot{\dot{a}}
\def\addot{\ddot{a}}

\def\bold#1{\setbox0=\hbox{$#1$}%
     \kern-.025em\copy0\kern-\wd0
     \kern.05em\copy0\kern-\wd0
     \kern-.025em\raise.0433em\box0 }

\def\GENITEM#1;#2{\par\vskip6pt \hangafter=0 \hangindent=#1
   \Textindent{$ #2$ }\ignorespaces}

\newcount\hour
\newcount\minute
\newtoks\amorpm
\hour=\time\divide\hour by60
\minute=\time{\multiply\hour by60 \global\advance\minute by-
\hour}
\edef\standardtime{{\ifnum\hour<12 \global\amorpm={am}%
    \else\global\amorpm={pm}\advance\hour by-12 \fi
    \ifnum\hour=0 \hour=12 \fi
    \number\hour:\ifnum\minute<100\fi\number\minute\the\amorpm}}
\edef\militarytime{\number\hour:\ifnum\minute<100\fi\number\minute}
\def\draftlabel#1{{\@bsphack\if@filesw {\let\thepage\relax
  \xdef\@gtempa{\write\@auxout{\string
    \newlabel{#1}{{\@currentlabel}{\thepage}}}}}\@gtempa
    \if@nobreak \ifvmode\nobreak\fi\fi\fi\@esphack}
     \gdef\@eqnlabel{#1}}
\def\@eqnlabel{}
\def\@vacuum{}
\def\draftmarginnote#1{\marginpar{\raggedright\scriptsize\tt#1}}
\def\draft{\oddsidemargin -.5truein
        \def\@oddfoot{\sl preliminary draft \hfil
        \rm\thepage\hfil\sl\today\quad\militarytime}
        \let\@evenfoot\@oddfoot \overfullrule 3pt
        \let\label=\draftlabel
        \let\marginnote=\draftmarginnote

\def\@eqnnum{(\theequation)\rlap{\kern\marginparsep\tt\@eqnlabel}%
\global\let\@eqnlabel\@vacuum}  }
\def\preprint{\twocolumn\sloppy\flushbottom\parindent 1em
        \leftmargini 2em\leftmarginv .5em\leftmarginvi .5em
        \oddsidemargin -.5in    \evensidemargin -.5in
        \columnsep 15mm \footheight 0pt
        \textwidth 250mmin      \topmargin  -.4in
        \headheight 12pt \topskip .4in
        \textheight 175mm
        \footskip 0pt

\def\@oddhead{\thepage\hfil\addtocounter{page}{1}\thepage}
        \let\@evenhead\@oddhead \def\@oddfoot{} \def\@evenfoot{}
}
\def\titlepage{\@restonecolfalse\if@twocolumn\@restonecoltrue\o
necolumn
     \else \newpage \fi \thispagestyle{empty}\c@page\z@
        \def\thefootnote{\fnsymbol{footnote}} }
\def\endtitlepage{\if@restonecol\twocolumn \else  \fi
        \def\thefootnote{\arabic{footnote}}
        \setcounter{footnote}{0}}  
\catcode`@=12
\relax
\def\be{\begin{equation}}
\def\ee{\end{equation}}
\def\bea{\begin{eqnarray}}
\def\eea{\end{eqnarray}}
\def\simlt{\stackrel{<}{{}_\sim}}

\def\mst11{m_{\;\widetilde{t}_{1}}}

\def\mst22{m_{\;\widetilde{t}_{2}}}
\def\mst12{m_{\;\widetilde{t}_{1,2}}}

\def\msb11{m_{\;\widetilde{b}_{1}}}
\def\msb22{m_{\;\widetilde{b}_{2}}}
\def\msb12{m_{\;\widetilde{b}_{1,2}}}

\def\mwidetilde2{\widetilde{m}^{2}}

\relax

\begin{document}
\input epsf

\topmargin-2.5cm
%
\begin{titlepage}
\begin{flushright}
CERN-TH/99-336\\
hep--ph/9911302 \\
\end{flushright}
\vskip 0.1in
\begin{center}
{\Large\bf Thermal and Non-Thermal Production of }
\vskip 0.2cm 
{\Large\bf Gravitinos in the Early Universe}

\vskip .5in
{\large\bf G.F. Giudice$^{1,}$\footnote{\baselineskip=16pt 
E-mail: {\tt Gian.Giudice@cern.ch }}}, {\large \bf A. Riotto$^{1,}$\footnote{\baselineskip=16pt Email:
{\tt riotto@nxth04.cern.ch}}} {\bf and}
{\large\bf I. Tkachev$^{2,}$\footnote{\baselineskip=16pt 
E-mail: {\tt Igor.Tkachev@cern.ch }}}

\vskip0.7cm
$^{1}$CERN Theory Division,

\vskip 0.2cm

CH-1211 Geneva 23, Switzerland.

\vskip 0.5cm

$^{2}$Institut f\"{u}r Theoretische Physik,                                           
\vskip 0.2cm
ETH-H\"{o}nggerberg, CH-8093, Z\"{u}rich, Switzerland.

\end{center}
\vskip 1cm
\begin{center}
{\bf Abstract}
\end{center}
\begin{quote}

The excessive production of gravitinos in the early universe destroys the successful predictions of nucleosynthesis. The thermal generation of gravitinos
after inflation leads to the  bound on the 
reheating temperature, $T_{RH}\simlt 10^9$ GeV. However, 
it has been recently realized that the non-thermal generation of gravitinos in the early universe can be extremely efficient and overcome the thermal production by several orders of magnitude, leading to much tighter constraints on the reheating temperature. In this paper, we first investigate
some  aspects of the  thermal production of gravitinos,  taking  into account that in fact  reheating is not   instantaneous and  inflation is likely to be followed by a prolonged stage of coherent oscillations of the inflaton field. We then proceed by further investigating the non-thermal generation of gravitinos, providing 
the  necessary tools to study this process in a generic time-dependent background with any   number of superfields. We also present the first numerical results regarding the non-thermal generation
of  gravitinos in particular supersymmetric models.

\end{quote}
\vskip1.cm
\begin{flushleft}
November  1999 \\
\end{flushleft}

\end{titlepage}
\setcounter{footnote}{0}
\setcounter{page}{0}
\newpage
%
\baselineskip=18pt
\noindent

\section{Introduction}

The overproduction of gravitinos represents a major obstacle in constructing  cosmological models based on supergravity \cite{sugra}.  Gravitinos decay very late and -- if  they 
are copiously produced during the evolution of the early universe --  their  
  energetic decay products    destroy the $^4$He and 
D nuclei by photodissociation, 
 thus jeopardizing the  successful nucleosynthesis predictions \cite{nucleo,ellis}. As a consequence, the ratio of the number density of gravitinos $n_{3/2}$ to the entropy density $s$ should be smaller than about $10^{-12}$ \cite{kaw} for gravitinos with mass of the order of 100 GeV. 

Gravitinos  can be produced in the early universe  because
of
thermal scatterings in the plasma during the stage of reheating after inflation. To avoid the overproduction of gravitinos one has to require that the reheating temperature $T_{RH}$ after inflation is not larger  than $\sim  (10^{8}-10^{9})\: {\rm GeV}$ \cite{ellis}. We will come back to this point and present a  detailed analysis of the thermal generation of gravitinos during reheating.

However, it has been recently realized that the  non-thermal effects occuring right after 
inflation because of the rapid oscillations of the inflaton field(s) 
provide an extra and very efficient source  of gravitinos \cite{linde,noi}. 
The
helicity $\pm 3/2$ part of the gravitino is excited only in tiny amounts,   
as the resulting abundance is always proportional to the  gravitino mass $m_{3/2}$ \cite{porc}. On the contrary, 
the helicity $\pm 1/2$  part obeys the equation of motion of a normal helicity $\pm 1/2$  
Dirac particle
in a background whose frequency is a combination of the different mass  
scales
at hand: the rapidly varying  superpotential mass parameter
of the fermionic superpartner of the scalar field  
whose $F$-term breaks supersymmetry, the Hubble rate and the gravitino  
mass \cite{linde,noi}. The non-thermal   
production of helicity $\pm 1/2$ gravitinos turns out to be much more efficient than their
thermal generation during the reheat stage after  
inflation \cite{linde,noi} and it was claimed  that the ratio $n_{3/2}/s$ for helicity $\pm 1/2$
gravitinos in generic supersymmetric models of inflation  
is roughly given 
by $10^{-2}T_{RH}/V^{1/4}$, where $V^{1/4}\sim 10^{15}$ GeV is the height of  
the potential during inflation. This 
leads to a very tight   upper bound on the reheat temperature, $T_{RH}\simlt 10^5 (V^{1/4}/10^{15}\:{\rm GeV})$ GeV \cite{noi}.

The production of   the helicity $\pm 1/2$ gravitino has been studied  in refs. \cite{linde,noi} starting from the supergravity Lagrangian and  in the simplest  case in which the energy density and the pressure of the universe are dominated by an oscillating  scalar field $\Phi$ belonging to a single
chiral superfield with minimal kinetic term. An application in the context of supersymmetric new inflation models has been recently presented in ref. \cite{new}.

 The  
 equation  describing the production of helicity $\pm 1/2$ gravitinos 
in supergravity reduces, in the limit in which the  amplitude of the oscillating field is smaller than the Planck scale, 
 to the equation describing the time evolution of 
the helicity $\pm 1/2$  Goldstino in global supersymmetry. This identification  explains why there is no 
suppression by inverse powers of $\mpl$ in the final number density of helicity $\pm 1/2$ gravitinos and is a simple manifestation of the gravitino-Goldstino equivalence theorem: on-shell
amplitudes with external helicity $\pm 1/2$ gravitinos are asymptotically  equivalent to amplitudes with corresponding external Goldstinos for energies
much larger than the gravitino mass \cite{eq}. This is analogous to the 
longitudinal $W$ bosons in the standard electroweak model behaving as Nambu-Goldstone bosons in the  high energy limit.

This simple observation  about the gravitino-Goldstino equivalence becomes  crucial   when the problem of computing the abundance of gravitinos generated by non-thermal effects  involves   more than one chiral superfield.  Describing the  production of  helicity $\pm 1/2$ gravitinos through  the 
equation of motion of the corresponding Goldstino in global supersymmetry is expected to  provide the correct result 
in the case in which the scalar  fields after inflation oscillate with amplitudes and frequencies smaller than the Planckian scale. Luckily, this situation is realized   in most of the  realistic supersymmetric models of inflation \cite{lr}.

The goal of this paper is twofold. In the first part of this work we will still concern ourselves with some aspects of the  thermal production of gravitinos during the reheating stage after inflation. We will perform a detailed analysis of such a process,  taking  into account the fact that reheating is far from being   instantaneous. Inflation is followed by a prolonged stage of coherent oscillations of the inflaton field.  
In this regime, the inflaton  is   decaying, 
but the inflaton energy has not yet been entirely converted into radiation. The temperature  $T$ rapidly increases to a maximum value and then slowly decreases as $a^{-3/8}$, being $a$ the scale factor of the universe. 
Only when the decay rate of the inflaton becomes of the order of the Hubble rate,  the universe enters the radiation-dominated phase and  one can properly define the reheat temperature $T_{RH}$. During this complicated dynamics, both gravitinos and  entropy are continously  generated  and one has to solve a set of Boltzmann equations to compute the final   ratio  $n_{3/2}/s$.  

In the second part of this work we will be dealing with the non-thermal production of gravitinos during the preheating stage after inflation.

Our aim is to provide the reader with all the tools necessary to study the helicity $\pm 1/2$ gravitino production in a generic time-dependent background and with a generic number of superfields. 
To achieve this goal, we will derive the master  equation of motion of the Goldstino in global supersymmetry with a generic number of superfields
and show that, in the case of one single chiral superfield and amplitudes
of the oscillating fields smaller than the Planck scale, it exactly reproduces the equation of motion of the helicity $\pm 1/2$ gravitino found in refs. \cite{linde,noi} starting from the supergravity Lagrangian. As a special case, we will concentrate on the case of two chiral superfields, which is particularly relevant when dealing with supersymmetric models of hybrid inflation. We will also present  the first complete numerical computation of the number density of the helicity $\pm 1/2$ gravitinos during the stage of preheating  for one single chiral superfield. This numerical  analysis will be performed  keeping all the   supergravity structure of the theory.

The paper is organized as follows. In sect.~2 we 
comment about the thermal production of gravitinos. In sect. ~3, we show how to derive
the equation of motion of the Goldstino in a generic time-dependent background,
we reproduce the helicity $\pm 1/2$ gravitino equation found in supergravity for one single chiral superfield, we comment upon   the decay rate of the helicity $\pm 1/2$ gravitinos and present the numerical results regarding
the number density of gravitinos in particular supersymmetric models
containing a single chiral superfield. Finally, in sect.~4 we discuss   the non-thermal production of gravitinos for the case of two chiral superfields,  which is relevant for realistic supersymmetric models of inflation.

\section{Aspects of thermal production of gravitinos  during   reheating}

At the end of inflation  the energy density of the
universe is locked up in a combination of kinetic energy and potential
energy of the inflaton field, with the bulk of the inflaton energy
density in the zero-momentum mode of the field.  Thus, the universe at
the end of inflation is in a cold, low-entropy state with few degrees
of freedom, very much unlike the present hot, high-entropy universe.
After inflation the frozen inflaton-dominated universe must somehow be
defrosted and become a high-entropy radiation-dominated universe.

The process by which the inflaton energy density is converted to
radiation is known as ``reheating'' \cite{book}.  The reader should rememeber that -- even if the process of reheating is anticipated by a stage of preheating \cite{kls} --  the efficiency of
preheating is very sensitive to the model and the model parameters.
In some models the process is inefficient; in some models it is not
operative at all.  Even if preheating is relatively efficient, it is
unlikely that it removes   {\it all} of the energy density  of the inflaton
field.  In particular, already during the resonant decay of the inflaton field, back-reaction processes of rescattering \cite{kt1} always create a sizeable
population of inflaton quanta with non-zero momentum \cite{kt2} which do not partecipate in the resonant decay. 
It is therefore likely that a stage during which the inflaton field is slowly decaying  is
necessary to extract the remaining inflaton field energy. This is exactly the stage 
we are going to analyze in this section.

The simplest way to envision this process is if the comoving energy
density in the zero mode of the inflaton (or the soft quanta generated in the process of rescattering during preheating)
decays into normal particles,
which then scatter and thermalize to form a thermal background.  It is
usually assumed that the decay width of this process is the same as
the decay width of a free inflaton field.

There are two reasons to suspect that the inflaton decay width might
be small.  The requisite flatness of the inflaton potential suggests a
weak coupling of the inflaton field to other fields since the
potential is renormalized by the inflaton coupling to other fields.  However, this restriction may be evaded in
supersymmetric theories where the nonrenormalization theorem ensures a
cancelation between fields and their superpartners.  A second and  basic  reason
to suspect weak coupling is that in local supersymmetric theories
gravitinos are produced during reheating.  Unless reheating is
delayed, gravitinos will be overproduced, leading to a large undesired
entropy production when they decay after big-bang nucleosynthesis.

As we already mentioned, of particular interest is a quantity known as the reheat temperature,
denoted as $T_{RH}$. In the oversimplified treatment, the reheat temperature is calculated by assuming
an instantaneous conversion of the energy density in the inflaton
field $\phi$ into radiation when the decay width of the inflaton energy,
$\Gamma_\phi$, is equal to $H$, the expansion rate of the universe.

The reheat temperature is calculated quite easily \cite{book}.  After
inflation the inflaton field $\phi$ executes coherent oscillations about the
minimum of the potential.  Averaged over several oscillations, the
coherent oscillation energy density redshifts as matter: $\rho_\phi
\propto a^{-3}$, where $a$ is the Robertson--Walker scale factor.  If
we denote as $\rho_I$ and $a_I$ the total inflaton energy density and
the scale factor at the initiation of coherent oscillations, then the
Hubble expansion rate as a function of $a$ is ($M_{Pl}=\sqrt{8\pi}\mpl$ is the Planck mass)
\begin{equation}
H^2(a) = \frac{8\pi}{3}\frac{\rho_I}{M^2_{Pl}}
        \left( \frac{a_I}{a} \right)^3.
\end{equation}
Equating $H(a)$ and $\Gamma_\phi$ leads to an expression for $a_I/a$.
Now if we assume that all available coherent energy density is
{\it instantaneously} converted into radiation at this value of $a_I/a$, we
can define the reheat temperature by setting the coherent energy
density, $\rho_\phi=\rho_I(a_I/a)^3$, equal to the radiation energy
density, $\rho_R=(\pi^2/30)g_*T_{RH}^4$, where $g_*$ is the effective
number of relativistic degrees of freedom at temperature $T_{RH}$.
The result is
\begin{equation}
\label{eq:TRH}
T_{RH} = \left( \frac{90}{8\pi^3g_*} \right)^{1/4}
                \sqrt{ \Gamma_\phi M_{Pl} } \
       = 0.2 \left(\frac{200}{g_*}\right)^{1/4}
              \sqrt{ \Gamma_\phi M_{Pl} } \ .
\label{eq:trh2}
\end{equation}

\subsection{Thermal production of dangerous relics in the case of instantaneous reheating}

Under the approximation of instantaneous reheating, the number density of any dangerous gravitational relic $X$ is readily solved. The Boltzmann equation reads
\begin{equation}
\label{l}
\frac{d n_{X}}{dt}+3H n_{X}\simeq\langle \sigma_{{\rm tot}} |v|\rangle n^2_{{\rm light}},
\ee
where $\sigma_{{\rm tot}}\propto 1/\mpl^2$ is the total cross section determining the rate of production of the gravitational relic and $n_{{\rm light}}\sim T^3$ represents the number density of light particles in the thermal bath. 

Since  thermalization is by hypothesis  very fast, the friction term $3H n_{X}$ in Eq. (\ref{l}) can be neglected and using the fact that the universe is radiation-dominated, {\it i.e.} $H\sim t^{-1}\sim T^2/\mpl$, one finds 

\beq
n_X\propto \frac{T^4}{\mpl}.
\eeq  
The  number
 density at thermalization in units of entropy density   reads
\be
\label{ll}
\frac{n_{X}}{s}\simeq 10^{-2}\:\frac{T_{RH}}{\mpl}.
\ee
As mentioned in the introduction, the slow decay rate of the $X$-particles
is the essential source of the cosmological problems because the  decay products of the gravitational relics  will destroy the $^4$He and D nuclei by photodissociation, 
and thus successful nucleosynthesis predictions. The most stringent bound comes from the resulting 
overproduction of D $+$ $^3$He, which would require that the relic abundance is smaller than $\sim 10^{-12}$ relative to the entropy density at the time of reheating after inflation \cite{kaw}
\be
\label{lll}
\frac{n_{X}}{s}\simlt 10^{-12}.
\ee

Comparing Eqs. (\ref{lll}) and (\ref{ll}), one may obtain an upper bound on the reheating temperature after inflation \cite{ellis} 
\be
T_{RH}\simlt (10^{8}-10^{9})\: {\rm GeV}.
\ee
If $T_{RH}\sim M_{{\rm GUT}}$, dangerous relics such as gravitinos would be abundant during nucleosynthesis and destroy the good agreement of the theory with observations. However, if  the reheating temperature satisfies the gravitino bound,  it is   too low to create superheavy GUT bosons that eventually decay and produce the baryon asymmetry \cite{review}.

In  the discussion above, the  crucial quantity   which determines the  abundance of dangerous relics after reheating is  the reheat temperature $T_{RH}$ (or the inflaton decay rate  $\Gamma_\phi$ through Eq. (\ref{eq:TRH})). 
 The reheat temperature is calculated by assuming
an instantaneous conversion of the energy density in the inflaton
field into radiation when the decay width of the inflaton energy
is equal to the  the expansion rate of the universe.

However,  the reheating process is not instantaneous. Right after inflation the decay width of the inflaton is expected to be much smaller than the 
Hubble rate,  $\Gamma_\phi\ll H$, otherwise $T_{RH}$ will violate the gravitino bound. Therefore, the universe undergoes a very long  period 
of matter-domination during which the energy density is dominated by the 
oscillations of the inflaton field  around the minimum of its potential. These oscillations last till the cosmic time becomes of the order of the lifetime of the inflaton field.  

In this
early-time and prolonged  regime of inflaton oscillations, the inflaton  is nevertheless  decaying, $\rho_\phi\propto e^{-\Gamma_\phi t}$, 
but the inflaton energy has not yet been entirely converted into radiation. The temperature  $T$ has the following behaviour. 
When the inflaton oscillations start and a small portion of the inflaton energy density has been transferred to radiation, the temperature rapidly grows to reach a maximum value $T_{MAX}$ and then it decreases scaling 
 as $a^{-3/8}$, which implies that the entropy per comoving volume $S$
is created: $S\propto a^{15/8}$ \cite{turner,book}. During this long   stage, the universe is not yet radiation-dominated. Finally, when $t\sim \Gamma_\phi^{-1}$, the inflaton energy density gets converted entirely into radiation and the universe enters the radiation-dominated phase. Only at this point one can properly define the reheat temperature $T_{RH}$. Indeed, the reheat temperature  is  best regarded as the temperature below which
the universe expands as a radiation-dominated universe, with the scale
factor decreasing as $g_*^{-1/3}T^{-1}$, where $g_*$ is the number of
relativistic degrees of freedom.  In this regard it has a
limited meaning \cite{turner,book}.

When studying the production of dangerous relics  during reheating,  it is necessary to take
into account the fact that reheating is not instantaneous and that 
the maximum temperature is greater than
$T_{RH}$. This implies that $T_{RH}$ {\it
should not} be used as the maximum temperature obtained in the
universe during reheating.  The maximum temperature is, in fact, much
larger than $T_{RH}$\footnote{As an application of this, particles of mass as large as $2\times 10^3$ times the reheat temperature $T_{RH}$ may be produced in interesting abundance to serve as dark-matter candidates \cite{ckr}.} and it is  inconsistent
to solve the Boltzmann equation for the gravitational relics  assuming that throughout the period of reheating $n_{{\rm light}}\sim T^3\propto a^{-3}$ and that 
the reheat temperature $T_{RH}$ is the largest temperature of the thermal system after inflation.  The goal of the next subsection is to provide a more appropriate computation of the number density of dangerous relics generated during the process of reheating.  
For sake of simplicity, we will focus on the gravitino case, but our results may be easily extended to other dangerous gravitational relics.

\subsection{A more appropriate approach to thermal generation of gravitinos during  reheating}

Let us consider a model universe with three components: inflaton field
energy, $\rho_\phi$, radiation energy density, $\rho_R$, and the
number density of the gravitino, $n_{3/2}$.  We
will assume that the decay rate of the inflaton field energy density
into radiation is $\Gamma_\phi$.   We will also assume that the light degrees
of freedom are in local thermodynamic equilibrium.  This is by no
means guaranteed, but the  analysis performed in ref. \cite{ckr}  shows that,  even if
thermalization does not occur, production of gravitinos during
reheating is not much different.

With the above assumptions, the Boltzmann equations describing the
redshift and interchange in the energy density among the different
components is
\begin{eqnarray}
\label{eq:BOLTZMANN}
& &\dot{\rho}_\phi + 3H\rho_\phi +\Gamma_\phi\rho_\phi = 0,
        \nonumber \\
& & \dot{\rho}_R + 4H\rho_R - \Gamma_\phi\rho_\phi= 0, \nonumber \\
& & \dot{n}_{3/2} + 3H n_{3/2} 
    + \langle \sigma_{{\rm tot}}|v|\rangle
        \left[ n_{3/2}^2 - \left( n_{3/2}^{EQ} \right)^2 \right]
        = 0,
\end{eqnarray}
where dot denotes time derivative. Here  $\langle
\sigma_{{\rm tot}}|v| \rangle$ is the total thermal average of the cross section times the M{\o}ller flux factor giving rise to the gravitino production and we have neglected the back-reaction of the gravitino abundance on the radiation energy density. 
 The equilibrium energy
density for the gravitinos, $n_{3/2}^{EQ}$, is determined by the
radiation temperature, $T$.

It is useful to introduce two dimensionless constants, $\alpha_\phi$
and $\alpha_X$, defined in terms of $\Gamma_\phi$ and $\langle \sigma
|v| \rangle$ as
\begin{equation}
\Gamma_\phi = \alpha_\phi M_\phi \qquad
\langle \sigma |v| \rangle = \alpha_X m_{3/2}^{-2} \ .
\end{equation}
For a reheat temperature much smaller than $M_\phi$, $\Gamma_\phi$
must be small.  From Eq.\ (\ref{eq:TRH}), the reheat temperature in
terms of $\alpha_\phi$ and $M_\phi$ is $T_{RH}\simeq \alpha_\phi^{1/2}
\sqrt{M_\phi M_{Pl}}$.  For $M_\phi=10^{13}$ GeV, $\alpha_\phi$ must be
approximately smaller than  $10^{-13}$.  

 It is also convenient to work with rescaled
quantities that can absorb the effect of expansion of the universe.
This may be accomplished with the definitions
\begin{equation}
\label{def}
\Phi \equiv \rho_\phi M_\phi^{-1} a^3 \ ; \quad
R    \equiv \rho_R a^4 \ ; \quad
X    \equiv n_{3/2} a^3 \ .
\end{equation}
It is also convenient to use the scale factor, rather than time, for
the independent variable, so we define a variable $x = a M_\phi$.
With this choice the system of equations (\ref{eq:BOLTZMANN}) can be written as (prime
denotes $d/dx$)
\begin{eqnarray}
\label{eq:SYS}
\Phi' & = & - c_1 \ \frac{x}{\sqrt{\Phi x + R}}   \ \Phi, \nonumber \\
R'    & = &   c_1 \ \frac{x^2}{\sqrt{\Phi x + R}} \ \Phi, 
             \nonumber \\
X'    & = & - c_3 \ \frac{x^{-2}}{\sqrt{\Phi x +R}} \
                \left( X^2 - X_{EQ}^2 \right).
\end{eqnarray}
The constants $c_1$, $c_2$, and $c_3$ are given by
\begin{equation}
c_1 = \sqrt{\frac{3}{8\pi}} \frac{M_{Pl}}{M_\phi}\alpha_\phi \ \qquad
c_2 = c_1\frac{M_\phi}{M_X}\frac{\alpha_X}{\alpha_\phi} \ \qquad
c_3 = \sqrt{\frac{3}{8\pi}}\alpha_X\frac{M_{Pl}M_\phi}{m_{3/2}^2} \ .
\end{equation}
$X_{EQ}$ is the equilibrium value of $X$, given in terms of the
temperature $T$ as \begin{equation}
X_{EQ} = \frac{3}{4}\frac{\zeta(3)}{\pi^2}g_{3/2}
        x^3 \left(\frac{T}{M_\phi}\right)^{3} \ .
\end{equation}
The temperature depends upon $R$ and $g_*$, the effective number of
degrees of freedom in the radiation:
\begin{equation}
T = \left( \frac{30}{g_*\pi^2}\right)^{1/4}
M_\phi \frac{R^{1/4}}{x} \ .
\end{equation}

It is straightforward to solve the system of equations in Eq.\
(\ref{eq:SYS}) with initial conditions at $x=x_I$ of $R(x_I)=X(x_I)=0$
and $\Phi(x_I)=\Phi_I$.  It is convenient to express
$\rho_\phi(x=x_I)$ in terms of the expansion rate at $x_I$, which
leads to
\begin{equation}
\Phi_I = \frac{3}{8\pi} \frac{M^2_{Pl}}{M_\phi^2}
                \frac{H_I^2}{M_\phi^2}\ x_I^3 \ .
\end{equation}
Before numerically solving the system of equations, it is useful to
consider the early-time solution for $R$.  Here, by early time, we
mean $H \gg \Gamma_\phi$, i.e., before a significant fraction of the
comoving coherent energy density is converted to radiation.  At early
times $\Phi \simeq \Phi_I$, and $R\simeq X \simeq 0$, so the equation
for $R'$ becomes $R' = c_1 x^{3/2} \Phi_I^{1/2}$.  Thus, the early
time solution for $T$ is simple to obtain \cite{ckr}
\begin{equation}
\label{threeeights}
\frac{T}{M_\phi} \simeq \left(\frac{12}{\pi^2g_*}\right)^{1/4}
c_1^{1/4}\left(\frac{\Phi_I}{x_I^3}\right)^{1/8}
        \left[ \left(\frac{x}{x_I}\right)^{-3/2} -
                \left(\frac{x}{x_I}\right)^{-4} \right]^{1/4}
                \qquad (H \gg \Gamma_\phi) \ .
\label{eq:approxtovmphi}
\end{equation}
Thus, $T$ has a maximum value of
\begin{eqnarray}
\frac{T_{MAX}}{M_\phi}& = & 0.77
   \left(\frac{12}{\pi^2g_*}\right)^{1/4} c_1^{1/4}
   \left(\frac{\Phi_I}{x_I^3}\right)^{1/8} \nonumber \\ & = & 0.77
   \alpha_\phi^{1/4}\left(\frac{9}{2\pi^3g_*}\right)^{1/4} \left(
   \frac{M_{Pl}^2H_I}{M_\phi^3}\right)^{1/4} \ ,
\end{eqnarray}
which is obtained at $x/x_I = (8/3)^{2/5} = 1.48$.  It is also
possible to express $\alpha_\phi$ in terms of $T_{RH}$ and obtain
\begin{equation}
\label{max}
\frac{T_{MAX}}{T_{RH}} = 0.77 \left(\frac{9}{5\pi^3g_*}\right)^{1/8}
                \left(\frac{H_I M_{Pl}}{T_{RH}^2}\right)^{1/4} \ .
\end{equation}

For an illustration, in the simplest model of chaotic inflation $H_I
\sim M_\phi$ with $M_\phi \simeq 10^{13}$ GeV, which leads to
$T_{MAX}/T_{RH} \sim 10^3 (200/g_*)^{1/8}$ for $T_{RH} =
10^9$ GeV.

For $x/x_I>1$, in the
early-time regime $T$ scales as $a^{-3/8}$, which implies that entropy
is created in the early-time regime \cite{turner}.  So if one is
producing gravitinos during reheating it is necessary to take
into account the fact that the maximum temperature is greater than
$T_{RH}$, and that during the early-time evolution, $T\propto
a^{-3/8}$.

The equation of motion of the number density of the gravitino is easily solved numerically. The results are plotted in Fig. 1. The total cross section for the gravitino production is such that $\alpha_X\simeq 16.6 (m_{3/2}/\mpl)^2$ while the total number of relativistic degrees of freedom is $g_*\simeq 230$. The inflaton parameters have been chosen to have $T_{RH}=10^9$ GeV, which gives $T_{MAX}\simeq 10^{12}$ GeV.

 \begin{figure}
\centering
\leavevmode\epsfysize=3.2in \epsfbox{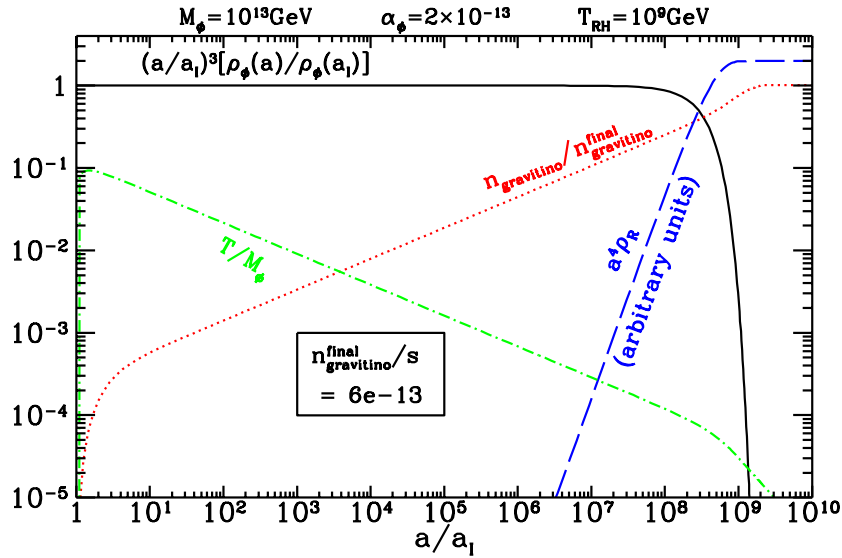}
\caption{ The time dependence of the temperature, the radiation energy density, the inflaton energy density and the gravitino number density for the case $T_{RH}=10^9$ GeV}
\label{fig:Fig1}
\end{figure}


We observe  that   the quantity $(n_{3/2}/s)/(n_{3/2}/s)_{{\rm final}}$, where $s$ is the entropy density, gradually increases with time  when $\Gamma_\phi$ is smaller than $H$, but remains always smaller than unity until the inflaton decays at $t\sim \Gamma_\phi^{-1}$. This means that most of the gravitinos  are  produced at the last stage of reheating when the inflaton decays and it makes sense to  talk about $T_{RH}$. We have also checked that $(n_{3/2}/s)_{{\rm final}}$ approximates  well     the usual estimate one gets neglecting the non-trivial evolution of the temperature of the radiation during the period $\Gamma_\phi\ll H$. This result can be explained   recalling that
 -- during the coherent oscillation epoch -- the entropy per comoving volume is increasing and the abundance of the just-produced gravitinos is continuously diluted by the entropy release. We have also checked that the final
  number density of gravitinos has 
a  dependence, even though weak, on the frequency of the inflaton oscillations  $M_\phi$. This dependence is not present in the case of instantaneous reheating,
where the number density of gravitinos depends only upon the reheating temperature and not on the frequency of the inflaton oscillations.

We conclude that -- even though the {\it maximum} temperature of $10^{12}$ GeV seems to be in contradiction with the usually quoted  upper bound of $(10^9-10^{10})$,    imposing the constraint
(\ref{lll}) gives  the usual upper bound on the reheating temperature 
$T_{RH}\simlt 10^9$ GeV. One should keep in mind, however, that the thermal evolution of the universe before the epoch $t\sim \Gamma_\phi^{-1}$ is  nonstandard and the physics leading to   the bound $T_{RH}\simlt 10^9$ GeV is much more involved than is 
 usually thought. This observation might be relevant when dealing  with either
a different parameter space for the gravitino, {\it e.g.} if the gravitino is very light,  or with  other kinds of dangerous relics.

\section{Non-thermal production of gravitinos and the gravitino-Goldstino  equivalence}

As shown in refs. \cite{linde,noi}, non-thermal effects occuring right after 
inflation due to  the rapid oscillations of the inflaton field(s) may lead to  copious 
gravitino production. As we noted in the introduction, this occurs because the helicity $\pm 1/2$ part of the 
gravitino can be efficiently excited during the evolution of the Universe after 
inflation. The non-thermal generation can be  extremely efficient and  
overcome the thermal production by several orders of magnitude,
in  realistic supersymmetric inflationary models.

The equation of the helicity $\pm 1/2$ gravitino has been found in refs. \cite{linde,noi} in the case in which the energy density and the pressure of the universe are dominated by an oscillating  scalar field $\Phi$ belonging to a single
chiral superfield with minimal kinetic term. 
In this section we would like to study the non-thermal production of gravitinos in 
a generic time-dependent gravitational background and for a generic number of chiral superfields.

The equation of the helicity $\pm 1/2$ gravitino with a  single
chiral superfield  and minimal kinetic term
is  identical to the
familiar equation for a spin-1/2 fermion in a 
time-varying background with frequency $\Omega$, which depends 
upon all the mass scales appearing in the 
problem, namely the Goldstino mass parameter
$\partial_\Phi^2 W$ (where $W(\Phi)$ denotes the superpotential),  the Hubble 
rate $H$ and the gravitino mass $m_{3/2}$. In the limit in which the amplitude of the oscillating  
scalar field is small,  
$|\Phi|\ll \mpl$, the frequency of the oscillations tends to 
$ \partial_\Phi^2 W$. The frequency $\Omega$ corresponds to the superpotential mass parameter
of the Goldstino which 
is `eaten' by the gravitino when supersymmetry is broken. Therefore, 
the equation  describing the production of helicity-1/2 gravitinos 
in supergravity reduces, in the limit of 
$|\Phi|\ll \mpl$, to the equation describing the time evolution of 
the helicity-1/2 Goldstino in global supersymmetry and no
suppression by powers of $\mpl$ is present.

This does not come as a surprise  and is in agreement with the gravitino-Goldstino equivalence theorem.
In spontaneously broken supergravity, the initially massless gravitino acquires a mass through
the superhiggs mechanism \cite{vol,cre}, by absorbing the Goldstino which
disappears from the physical spectrum. Before becoming massive, the gravitino, which is a Majorana spin $3/2$ particle, posseses only $\pm 3/2$ helicity states. The Goldstino, a Majorana fermion, provides for its missing (longitudinal) $\pm 1/2$ states. The equivalence theorem  is valid in the limit of large energies compared to $m_{3/2}$ where  the longitudinal component of the gravitino effectively behaves as a spin $1/2$ Goldstino  \cite{eq}. 
 
Therefore, it appears of advantage to compute the equation of motion of  the helicity $\pm 1/2$ gravitino by finding the
equation of motion of the corresponding Goldstino in global supersymmetry. This procedure is particularly welcome when the problem involves   more than one chiral superfield and is expected to  provide the correct result for the number density of helicity $\pm 1/2$ gravitinos
in the case in which the scalar  fields after inflation oscillate with amplitudes and frequencies smaller than the Planckian scale. This is exactly what is realized    in most of the  realistic supersymmetric models of inflation \cite{lr}.

\subsection{The equation of motion of the Goldstino in global supersymmetry and in a time-dependent background}

Let us now  find  the
equation of motion of the Goldstino when the energy density of the background
is dominated by a set of scalar fields following the trajectories imposed by their equation of motion. We will therefore suppose that the scalar fields are displaced from the minima of their potential and are free to oscillate about such minima. This is what happens right after inflation and during the preheating stage. 

The identification of the Goldstino requires a generalization of the standard procedure used in the static case, that is  when the scalar fields are 
sitting at the minima of their potential and the cosmological constant vanishes. In the following we will neglect the expansion of the universe. For the practical purpose of computing the number density of helicity $\pm 1/2$ gravitinos generated during the preheating stage, 
this is good approximation since the non-thermal production of gravitinos is expected to overcome the thermal generation in those supersymmetric models
in which the frequency of the oscillations of the scalar fields is much larger than the rate of the expansion of the universe and 
most of the gravitinos are generated  within the first few oscillations. Neglecting the expansion of the universe will also make the identification of the Goldstino more transparent. Finally,  we will not  concern ourselves with    a theory charged under some gauge group, but suppose that during the evolution of the system some $F$-term is nonvanishing. Our findings can be easily generalized to include the possibility
that supersymmetry is broken by some (time-dependent) $D$-term.

Consider a global supersymmetric theory with Lagrangian 
\begin{eqnarray}
&&{\cal L}=\partial_\mu z_i\partial^\mu z^i+\frac{i}{2}\bar{\chi}_i\gamma_\mu\partial^\mu \chi_i-V(z^i,z_i)-\frac{1}{2}\left(W_{ij} \bar{\chi}_i P_L\chi^j+{\rm h.c.}\right),
\nonumber\\
&&V(z^i,z_i)= W_i (W^{\dagger})^i.
\end{eqnarray}
Here $W=W(z^i)$ is the superpotential, $z^i$ and $\chi_i$ denote a set the scalar and fermionic fields respectively, $z_i=(z^i)^\dagger$, $W_i=\partial W/\partial z^i$ and $P_L=(1-\gamma_5)/2$ is the left-handed projection operator.
The index $i$ runs from 1 to $N$, being $N$ the number of multiplets and we use the standard convention that the  sum is intended when the index $i$ is contracted.

The Lagrangian is invariant under the following set of supersymmetric transformations 
\begin{eqnarray}
\label{p}
\delta_\varepsilon z^i&=&\sqrt{2}\bar{\varepsilon} P_L\chi^i,\nonumber\\
\delta_\varepsilon \chi^i&=&\sqrt{2}\Theta^i\varepsilon,
\end{eqnarray}
where $\varepsilon$ is the spinor parametrizing the infinitesimal supersymmetric transformation and we have defined the matrix
\begin{equation}
\label{a}
\Theta^i=\left[i \dslash z^i-  (W^{\dagger })^i\right].
\end{equation}

Given  a generic background, supersymmetry is broken if   
\begin{equation}
\langle \delta_\varepsilon \chi^i\rangle\neq 0.
\end{equation}
This happens if the expectation value of the matrix $\Theta^i$ is nonvanishing
\begin{equation}
\langle \Theta^i\rangle\neq 0.
\end{equation}
In particular, for a constant (time-independent) background, one recovers the usual condition that supersymmetry is broken if, for some field $z^i$, the $F$-term is nonvansishing   

\begin{equation}
\langle W_i\rangle\neq 0.
\end{equation}
On the other hand, in the  case of a time-dependent background, the breakdown of supersymmetry comes  also from the

\beq 
\gamma_0\frac{dz^i}{dt}
\eeq
piece in the matrix $\Theta^i$. This is not surprising   since supersymmetry is broken  in the early universe anytime  some form of nonvanishing energy density appears. This is what happens  during inflation and the subsequent stage of preheating and reheating when scalar fields oscillate around the minima of their potential.   

The Goldstone theorem tells us that the Goldstino is easily identified from the supersymmetric transformation (\ref{p})

\begin{equation}
\label{Goldstino}
\eta=\Theta_i\chi_i.
\end{equation}
We introduce now the two projection operators
\begin{eqnarray}
(P^\perp)_{ij}&=&\delta_{ij}-\frac{\Theta^{\dagger}_i}{\Theta^\dagger \Theta} \Theta_j,\nonumber\\
(P^{\parallel})_{ij}&=&\frac{\Theta^{\dagger}_i}{\Theta^\dagger \Theta} \Theta_j,
\end{eqnarray}
where we have defined $\Theta^\dagger \Theta=\Theta_k^{\dagger}\Theta_k$. The two operators project respectively onto the subspace orthogonal to the Goldstone fermion and onto the Goldstone fermion itself. 

Making use of the definition (\ref{a}), we find that, for a background of real fields, 
\begin{equation}
\Theta^\dagger \Theta=\sum_i\left[\left(\frac{dz^i}{dt}\right)^2+ \left(W_i\right)^2\right]=\rho.
\end{equation} 
Therefore, the combination $\Theta^\dagger \Theta$ gives the total energy density of the system $\rho$  which  -- if the expansion of the universe is neglected -- remains constant in time.

The spin $1/2$ field $\chi_i$ can be rewritten as
\begin{eqnarray}
\label{dec}
\chi_i&=&\chi_i^\perp+ \frac{\Theta^{\dagger}_i}{\Theta^\dagger \Theta}\eta,\nonumber\\
\chi_i^\perp&=& (P^\perp)_{ij}\chi_j.
\end{eqnarray}
Notice that  the fields $\chi_i^\perp$ are not linearly independent since they satisfy the following relation
\begin{equation}
\label{linear}
\Theta_i\chi_i^\perp=0.
\end{equation}
This condition tells us that one of the   $\chi_i^\perp$ fields may be expressed in terms of the remaining $(N-1)$ ones. 

We now choose the nonvanishing vacuum expectation values of the scalar fields in the real direction, ${\rm Re}\:z_i=\phi_i/\sqrt{2}$, ${\rm Im}\: z_i=0$. 
The equation of motion of the scalar and fermionic fields read
\begin{eqnarray}
\label{e1}
\ddot{\phi_i}&=&-W_{ij}W_j,\\
\label{e2}
i\gamma^0 \dot{\chi}_i&=&\hat{k}\chi_i+W_{ij}\chi_j,
\end{eqnarray}
where the dots stand for derivative with respect to  time, $\hat{k}=\vec{\gamma}\cdot\vec{k}$ and we have used the plane-wave ansatz $\chi_i\sim e^{i\vec{k}\cdot\vec{x}}$ for the space-dependent
part. The matrices $\Theta_i$ satisfy the following equation
\begin{equation}
\label{rel}
i\gamma^0 \dot{\Theta}_i=-W_{ij} \Theta_j.
\end{equation}
Inserting now the decomposition (\ref{dec}) into Eq. (\ref{e2}),  multiplying by $\Theta_i$ and $\Theta_i^\dagger$ respectively and making use of the Eq. (\ref{rel}), we get the following equations of motion
\begin{eqnarray}
\label{a1}
i\gamma^0\dot{\eta}&=&\hat{k}{\cal G}^\dagger\eta+\hat{k}\hat{\chi},\\
\label{a2}
i\gamma^0\dot{\hat{\chi}}&=&-i{\cal G}^\dagger\gamma^0\dot{\eta}+\hat{k}\eta
+2\Theta_i^\dagger W_{ij}\chi^{\perp}_j,
\end{eqnarray}
where we have defined the following combinations
\begin{eqnarray}
{\cal G}&=&\frac{\Theta_i \Theta_i}{\Theta^\dagger \Theta},\nonumber\\
\hat{\chi}&=&\Theta_i^\dagger \chi^{\perp}_i.
\end{eqnarray}
The matrix ${\cal G}$ can be expressed in terms of the energy density $\rho$ and the pressure $p$ of the oscillating scalar fields
\begin{equation}
{\cal G}=-\frac{p+2i\gamma^0\dot{W}}{\rho}.
\end{equation}
Differentiating Eq. (\ref{a1}) with respect to   time and using Eq. (\ref{a2}) we find the master equation of motion of  the Goldstino
\begin{equation}
\label{master}
\ddot{\eta}+k^2\eta +i\gamma^0\hat{k}\dot{{\cal G}}^\dagger\eta-2i\hat{k}\gamma^0
\dot{\Theta}_i^\dagger \chi^{\perp}_i=0.
\end{equation}
This equation is valid for a generic number of chiral superfields and -- because of the 
gravitino-Goldstino  equivalence theorem -- is expected to provide the necessary tool to describe the production of helicity $\pm 1/2$ gravitinos during the preheating stage after inflation, when the typical energy of the system  and field amplitudes   (or the frequencies of the oscillations of the scalar fields) are large compared to the gravitino mass $m_{3/2}$ and smaller than the Planck scale. We notice the non-trivial result that any time-dependent
function has disappeared from the $k^2\eta$ term; in the ultraviolet regime,  
Eq. (\ref{master}) is solved by plane-waves and 
particle production shuts off as one would expect from general arguments.

In a static background for which $\dot{\phi}_i=0$, we have $\Theta_i=\Theta_i^\dagger$, 
$\dot{G}^\dagger=\dot{\Theta}_i=0$ and $\hat{\chi}=0$ by virtue of Eq. (\ref{linear}). The Goldstino equation is solved by plane-waves and -- as expected -- no particle production takes place. 

Let us now consider the special case of one   single
chiral superfield $\Phi$ with minimal kinetic term. We have $N=1$,  the only physical degree of freedom is the Goldstino and 

\beq
\chi^\perp=\hat{\chi}=0.
\eeq 
Eq. (\ref{a1}) simplifies to 

\beq
\label{ppp}
\left(i\gamma^0\partial_0-\hat{k}{\cal G}^\dagger\right)\eta=0,
\eeq
where 
\beq
{\cal G}=\frac{\Theta_1}{\Theta_1^\dagger},
\eeq
and  the matrix $\Theta_1$ is given in Eq. (\ref{a}) for the case $i=1$.
Notice that the matrix ${\cal G}$ has manifestly absolute value equal to unity
\beq
\label{pppp}
\left|{\cal G}\right|^2={\cal G}^\dagger {\cal G}= \frac{\Theta_1^\dagger}{\Theta_1}\times
\frac{\Theta_1}{\Theta_1^\dagger}={\bf 1}.
\eeq
Therefore, it is possible to rewrite ${\cal G}$ in the following form
\beq 
{\cal G}=e^{2i\gamma^0\varphi},
\eeq
By making a field redifinition $\eta\rightarrow {\rm exp}(i\gamma^0\varphi)    
\eta$, the equation of motion of the Goldstino becomes
\beq
\label{sss}
i\gamma^0\dot{\eta}-\hat{k}\eta-m_{{\rm eff}}\eta=0,
\eeq
where 

\beq
m_{{\rm eff}}=\dot{\varphi}=\frac{\partial^2 W}{\partial \Phi^2}
\eeq
and we have used the fact that ${\cal G}$ satisfies the following differential equation
\beq
\frac{\dot{{\cal G}}}{{\cal G}}=2i\gamma^0  \frac{\partial^2 W}{\partial \Phi^2}.
\eeq
Eq. (\ref{sss}) is the equation of motion of a spin-1/2 fermion in a time-dependent background given by the oscillating mass $m_{{\rm eff}}$.
We now wish to show that the Eq. (\ref{sss}) found for one single chiral superfield  in the limit of global supersymmetry  correctly reproduces -- for  amplitudes of the 
scalar field $\Phi$ much smaller than the Planck scale -- the equation of motion of the helicity $\pm 1/2$ gravitino found
in refs. \cite{linde,noi} starting from a local supersymmetric theory, {\it i.e.}  supergravity. 

\subsection{Non-thermal production of gravitinos in the  case of one chiral superfield}

Let us first remind the reader of some of the basic results obtained in refs. \cite{linde,noi} regarding the equation of motion of the helicity $\pm 1/2$ gravitino  in the case of one 
single chiral superfield and minimal kinetic term. 

If we  start with the  supergravity Lagrangian,  the 
single chiral fermion $\chi$ -- which is the superpartner of the scalar component in the chiral supermultiplet $\Phi$ -- plays the role of the Goldstino and   can be gauged away to zero, so that 
no mixing between the gravitino $\psi_\mu$ and $\chi$ is present. Under these circumnstances, the equation of motion of the gravitino becomes
\beq
R^\mu \equiv \epsilon^{\mu \nu \rho \sigma}\gamma_5 \hgam_\nu {\cal 
D}_\rho \psi_\sigma =0.
\label{rarcur}
\eeq
Here ${\cal 
D}_\rho$ is the covariant derivative and greek letters denote space-time indices. The condition ${\cal D}\cdot R =0$ gives the following algebraic 
constraint 
\beq
\label{con}
\hat\gamma_0\psi^0=c\sum_{i=1}^3\hgam_i\psi^i,
\eeq
where the matrix $c$, in the limit of $|\Phi|\ll \mpl$, reduces to
\begin{equation}
c=\frac{p+2i\gamma^0 \dot{W}}{\rho}=-{\cal G}.
\nonumber
\end{equation} 
Two degrees of freedom may be eliminated using Eq. (\ref{con}).

We note  that the constraint (\ref{con}) may be recovered in the following alternative way. The mixing term in the supergravity Lagrangian 
between the gravitino and the Goldstino is of the form 

\beq
\bar{\chi}\hgam^\mu
\Theta_1^\dagger\psi_\mu.
\eeq
By using the definition (\ref{Goldstino}) $\eta=\Theta_1\chi$ the mixing term
becomes 
\begin{equation}
\frac{1}{\Theta_1^\dagger \Theta_1}\bar{\eta}\Theta_1\hgam^\mu \Theta_1^\dagger \psi_\mu.
\end{equation}
Choosing the gauge in which such a term vanishes is equivalent to require that
\beq
\Theta_1\hgam^\mu \Theta_1^\dagger \psi_\mu=0.
\label{cond}
\eeq
This condition gives $\hat\gamma_0\psi^0=-{\cal G}\sum_{i=1}^3\hgam_i\psi^i$, which coincides with the constraint (\ref{con})\footnote{The constarint (\ref{cond})
is  easily generalized to the case of many superfields $
\Theta_i\hgam^\mu \Theta_i^\dagger \psi_\mu=0$.}.

Because of the antisymmetric properties of the
Levi-Civita symbol, the equation $R^0=0$ 
does not contain time derivatives and provides  another
 algebraic constraint on the gravitino momentum modes. Such a constraint allows to remove two extra degrees of freedom and to define  two physical Majorana fermion states $\psi_{3/2}$ and $\psi_{1/2}$ which may be shown to    correspond to the $\pm 3/2$
and $\pm 1/2$ helicity states respectively, 
by explicitly constructing the helicity 
projectors  in the flat limit \cite{noi}. 
The Lagrangian may be diagonalized as \cite{noi} 
${\cal L}={\cal L}_{3/2}+{\cal L}_{1/2}$, where 
\begin{eqnarray}
\label{la}
{\cal L}_{3/2}&=&\bar{\psi}_{3/2}\left[ i\gamma^0\partial_0 +i\frac{5\adot}{2a}\gamma^0 -m_{3/2}a\right] \psi_{3/2},\nonumber\\
{\cal L}_{1/2}&=&\bar{\psi}_{1/2}\left[ i\gamma^0\partial_0 +i\frac{5\adot}{2a}\gamma^0 +m_{3/2}a
+\hat{k}G\right] \psi_{1/2},
\end{eqnarray}
where $a$ is the scale factor, $G=A+i\gamma^0 B$ and $A$ and $B$ are time-dependent functions \cite{noi}
\begin{eqnarray}
\label{ab}
A&=&\frac{1}{3\left( \frac{\adot^2}{a^4}+m_{3/2}^2\right)^2}\left[
2\frac{\addot}{a^3}\left(m_{3/2}^2-\frac{\adot^2}{a^4}\right) +
\frac{\adot^4}{a^8}-4m_{3/2}^2\frac{\adot^2}{a^4}+3m_{3/2}^4\right.\nonumber\\
&-&\left.4\frac{\adot}{a^3}
\dot{m}_{3/2} 
m_{3/2}\right], \label{eqqa}\\
 B&=&\frac{2m_{3/2}}{3\left( \frac{\adot^2}{a^4}+m_{3/2}^2\right)^2}\left[
2\frac{\addot \adot}{a^5}-\frac{\adot^3}{a^6}+3m_{3/2}^2\frac{\adot}{a^2}
+\frac{\dot{m}_{3/2}}{m_{3/2}a}\left(m_{3/2}^2-\frac{\adot^2}{a^4}\right) \right] . \label{eqqb}
\eea
They may be expressed in terms the pressure and the energy density of the scalar field $\Phi$. Here time is  conformal  and the line element is
$ds^2=a^2(\tau) (d\tau^2 - d{\vec x}^2)$.

The diagonal
time and space components of the Einstein equation become
\bea 
\frac{\adot^2}{a^4}&=& \frac{1}{3\mpl^2}\left[ V(\Phi)+
\left| \frac{d\Phi}{dt}\right|^2
\right] ,\label{ein1} \\
2\frac{\addot}{a^3}-\frac{\adot^2}{a^4}&=& \frac{1}{\mpl^2}\left[ V(\Phi)-
\left| \frac{d\Phi}{dt}\right|^2
\right] .\label{ein2} 
\eea
Using the expression for
the gravitino mass $m_{3/2}$ in terms
of the superpotential $W$,
\beq
m_{3/2}=e^{\frac{\Phi^\dagger \Phi}{2\mpl^2}}~\frac{|W(\Phi)|}{\mpl^2} ,
\eeq
we can write the scalar potential $V$ as
\beq
V=e^{\frac{\Phi^\dagger \Phi}{\mpl^2}}
\left[ \left| \partial_\Phi W +\frac{\Phi^\dagger W}{\mpl^2}
\right|^2 -\frac{3|W|^2}{\mpl^2} \right] = m_{3/2}^2 \mpl^2 \left[ \left|
\frac{\dot{m}_{3/2} \mpl}{am_{3/2} \frac{d\Phi}{dt}} \right|^2 -3 \right] . \label{scap}
\eeq
Replacing Eqs.~(\ref{ein1}) and (\ref{ein2}) in Eq. (\ref{scap}), one 
obtains \cite{noi}
\beq
\dot{m}_{3/2}^2 = -\frac{\addot^2}{a^4}+\frac{\addot}{a}\left( \frac{\adot^2}{a^4}
-3m_{3/2}^2\right) +2\frac{\adot^4}{a^6} +6\frac{\adot^2}{a^2}m_{3/2}^2 . \label{magrel}
\eeq
When this expression for $\dot{m}_{3/2}$ is used in Eqs.~(\ref{eqqa}) and 
(\ref{eqqb}), we obtain the ramarkable property \cite{linde,noi}
\beq
\left|G^\dagger G\right|=A^2+B^2=1.
\label{cazz}
\eeq

We are now in the position to show 
the gravitino-Goldstino equivalence explicitly. To do so, we    neglect the expansion of the universe and the gravitino mass $m_{3/2}$ and consider the limit 
$|\Phi|\ll \mpl$. The matrix $G$ has the following limit \cite{linde,noi}

\beq
G\stackrel{|\Phi|\ll \mpl}{\longrightarrow} \frac{p-2i\gamma^0\dot{W}}{\rho}=
-{\cal G}^\dagger.
\eeq
The equation of motion of the helicity $\pm 1/2$ gravitino therefore  reduces to
\beq
\left(i\gamma^0\partial_0 -\hat{k}{\cal G}^\dagger\right) \psi_{1/2}=0.
\eeq
This equation is exactly reproduced in the global supersymmetric limit by the equation of motion of the Goldstino (\ref{ppp}).
 
One can also use the gravitino-Goldstino  equivalence  to explain the remarkable property that the matrix $G$ 
has absolute value equal to unity, by making use of Eq. (\ref{pppp}). 

\subsection{Comments on the gravitino decay}

We bragged about achieving a large number density of helicity $\pm 1/2$ gravitinos from non-thermal effects and how this phenomenon is strictly related to the fact that $\mpl$ does not appear in the 
equation of motions, but then tacitly assumed  that the comoving number  of  helicity $\pm 1/2$ gravitinos at nucleosynthesis is the same one which may be generated during preheating. This issue deserves a closer look because  
one might think that  helicity $\pm 1/2$ gravitinos  promptly decay (or rapidly thermalize), thus not leading to a large undesired
entropy production when they decay after big-bang nucleosynthesis. 

However, this is not the case; helicity $\pm 1/2$ gravitinos do have a decay rate which is
suppressed by the gravitational coupling  $\mpl^{-2}$ and is therefore small. This can be easily understood in the following way. The helicity $\pm 1/2$ components of the gravitino field correspond to the Goldstino, which is 
derivatively coupled to the supercurrent.  Hence,  the total amplitude for the decay rate of the helicity $\pm 1/2$ gravitino has to be proportional to the
mass splitting within the supermultiplets. For example,  in the present vacuum, where we suppose supersymmetry is  broken by some $F$-term
with $F\sim m_{3/2} \mpl$,  the coupling between the helicity $\pm 1/2$ gravitino, a fermion $f$ and its superpartner $\widetilde{f}$ is proportional
to $(m^2_{\widetilde{f}}-m_f^2)/F$. Since $(m^2_{\widetilde{f}}-m_f^2)\sim m_{3/2}^2$, the coupling is suppressed by $m_{3/2}/\mpl^{-1}$. Similarly, the coupling
of the  helicity $\pm 1/2$ gravitino, with a gauge boson and a gaugino is proportional to $m_\lambda/F\sim \mpl^{-1}$, where $m_\lambda\sim m_{3/2}$ is the gaugino mass.
This is the reason why, when dealing with thermal production of gravitinos during reheating, the helicity $\pm 1/2$ and $\pm 3/2$ gravitinos are treated on the same ground and  have both $\mpl$-suppressed cross sections. 

Right after inflation and during the preheating stage, supersymmetry is badly broken by the energy density $\rho$  stored in the oscillating scalar fields and what measures the breaking of supersymmetry is not a simple   $F$-term like in the present vacuum, but the parameter $\sqrt{\Theta^\dagger \Theta}=\rho^{1/2}$.
The helicity $\pm 1/2$ gravitinos may decay into lighter fermions and sfermions through a coupling proportional to $\Delta m^2/\rho^{1/2}$, where $\Delta m^2$ is the mass-splitting in the  given light supermultiplet. As supersymmetry breaking is transmitted by the gravitational force, at the preheating stage
$\Delta m^2$ is at most of the order of $\rho/\mpl^2\sim H^2$. If the decay is kinematically allowed,  the decay rate of the 
helicity $\pm 1/2$ gravitinos into fermions and sfermions is   at most

\beq
\label{decay}
\Gamma\sim \frac{\rho}{\mpl^4}\Omega\sim \left(\frac{H}{\mpl}\right)\left(\frac{\Omega}{\mpl}\right)H\ll H,
\eeq
 where $\Omega$ is the  (decreasing) time-dependent frequency  of the oscillations of the scalar fields responsible for the non-thermal production of the helicity $\pm 1/2$ gravitinos. 
Similarly, in the case of decay into gauge bosons plus gauginos, the decay rate is $\sim (\Omega/\mpl)^2\Omega\ll H$. 

These estimates are  valid as long as the oscillating scalar fields dominate the energy density of the universe. The  ``composition" of the helicity $\pm 1/2$ gravitino through the Goldstino mixture (\ref{Goldstino}) changes with time. 
During the prolonged  stage of coherent oscillations, the main contribution 
to the helicity $\pm 1/2$ gravitino    comes from the fermionic  superpartners of 
the coherently oscillating scalar fields and the decay rate (\ref{decay}) applies.
This decay rate is   tiny  and  always smaller   than the rate of the expansion of the universe; the number density of the 
helicity $\pm 1/2$ gravitinos does not drop during this epoch.
At later stages, the main contribution to the helicity $\pm 1/2$ gravitino  is given by the fermionic  superpartners of the scalar fields
whose $F$-terms break supersymmetry in the present vacuum. This means that --
when the composition of the helicity $\pm 1/2$ gravitino changes  with time --  the decay rate will smoothly interpolate between (\ref{decay}) and  the more familiar rate  $\Gamma\sim m_{3/2}^3/\mpl^2$. 
As the decay rate remains smaller    than the Hubble rate till after the   nucleosynthesis epoch, the amount of gravitinos per comoving volume generated by non-thermal effects during the preheating stage   remains frozen till the age of the universe becomes of the order of $\mpl^2/m_{3/2}^3$. At this moment, gravitinos decay and their  decay products destroy the light element abundances unless $n_{3/2}/s$ is sufficiently small.

\subsection{Numerical results for the case of one chiral superfield}

In this subsection we wish to provide the first complete numerical computation of the number density of the helicity $\pm 1/2$ gravitinos during the stage of preheating after inflation in the case in which the energy density of the universe is dominated by a single oscillating scalar field. It is important to keep in mind that    a generic supersymmetric inflationary stage dominated by an 
$F$-term has the problem that the flatness of the potential is 
spoiled by supergravity corrections or, in other words,
the slow-roll parameter 
$\eta=\mpl^2 V^{\prime\prime}/V$ gets contributions of
 order unity \cite{lr}. In simple one chiral field models based on  
superpotentials of the type $W=M_\phi\Phi^2/2$ or $W=\sqrt{\lambda}\Phi^3/3$,   supergravity corrections make inflation 
impossible to start. To construct a model of inflation in the context
 of  supergravity, one must either invoke 
 accidental 
cancellations \cite{linderiotto}, or a period of inflation dominated by a
$D$-term \cite{dterm},  or some  particular properties based  on string theory
\cite{gel}. Nevertheless, we  are not interested here in the inflationary stage, but rather on the subsequent stage of preheating. During this period, it might be that the superpotential is well-approximated by a quadratic or cubic 
expression along the oscillating scalar field. 

The equation for the helicity $\pm 1/2$ gravitino in the supergravity approach
with one single chiral superfield has been reduced to a more  familiar second-order differental equation for a spin-1/2 fermion in a 
time-varying background in refs. \cite{linde,noi}. We wish to present here a slightly different derivation. Since the matrix $G$ has absolute value equal to unity, it is possible to rewrite it in the following form
\beq 
G=e^{2i\gamma^0\varphi},
\eeq
where $\varphi$ is a  phase depending upon the conformal time. By making a field redifinition $\psi_{1/2}\rightarrow a^{-5/2}{\rm exp}(-i\gamma^0\varphi)    
\psi_{1/2}$, the Lagrangian ${\cal L}_{1/2}$ simplifies to

\beq
{\cal L}_{1/2}=\bar{\psi}_{1/2}\left[i \dslash -m_{{\rm eff}}a \right] \psi_{1/2}.
\eeq
This is   the Lagrangian for a spin-1/2 fermion in a 
time-varying background with effective mass   

\beq
m_{{\rm eff}}=-\left[ m_{3/2}+\left(\frac{\dot{\varphi}}{a}\right)\right],
\eeq 
where $\varphi=-(i/2)\gamma^0 (\dot{G}/G)$ and $G=A+i\gamma^0B$ is given in eqs. (\ref{ab}).
One
can use as a guide  the recent results obtained in the theory of generation of
Dirac fermions during and after inflation \cite{ferm}. During inflation, since the mass
scales present in the effective mass $m_{{\rm eff}}$  are approximately constant in time,
one does not expect a significant production of gravitinos (the number density
can be at most $n_{3/2}\sim H_I^3$, where $H_I$ is the value of the Hubble rate
during inflation). However, in the evolution of the Universe subsequent to
inflation, a large amount of gravitinos may  be produced. During the inflaton
oscillations, the Fermi distribution function is rapidly saturated up to some
maximum value of the momentum $k$, {\it i.e} $n_k\simeq  1$ for $k\simlt
k_{{\rm max}}$ and it is zero otherwise.  The resulting number density is
therefore $n_{k}\sim k_{{\rm max}}^3$. The value of $k_{{\rm max}}$ is
expected to be roughly of the order of the inverse of the time-scale needed for
the change of the mass scales of the problem at hand.

The field $\psi_{1/2}$ can be as usual
expanded in terms of Fourier modes of the form
\begin{equation}
\label{eq:diractwo}
\psi_{1/2} = \int \frac{d^3\!k}{(2\pi)^{3/2}}\ e^{-i\vec{k}\cdot\vec{x}}
\, \sum_{r=\pm 1} \left[ u_r(k,\eta)a_r(k) + v_r(k,\eta)b^\dagger_r(-k)\right] \ ,
\end{equation}
where the summation is over spin and the conditions  $v_r(k) \equiv C\bar{u}^T_r(-k)$ and $a_r=b_r$ are imposed by the fact that the gravitino is a Majorana particle.
The canonical anticommutation relations imposed upon the creation and
annihilation operators may be used to normalize the spinors $u$ and
$v$.

Defining $u_r \equiv \left[u_+(\eta)\psi_r(k), r u_-(\eta)\psi_r(k)\right]^T$ and 
$v_r \equiv \left[r v_+(\eta)\psi_r(k),  v_-(\eta)\psi_r(k)\right]^T$, where 
$\psi_r(k)$ are the  two-component eigenvectors of the helicity operators,  
 and using a
representation where $\gamma^0 = {\rm diag}({\bf 1},-{\bf 1})$,
Eq.\ (\ref{eq:diractwo}) can be written as two uncoupled second-order
differential equations for $u_+$ and $u_-$:
\begin{equation}
\label{eq:waveequation}
\ddot{u}_\pm + \left[ \omega_k^2 \pm i (m_{{\rm eff}}a)^{\cdot}
 \right] u_\pm = 0 \ ,
\end{equation}
where, $\omega_k^2= k^2 + m_{{\rm eff}}^2a^2$. 
In order to calculate the number density, we must first diagonalize
the Hamiltonian.  In the basis of Eq.\ (\ref{eq:diractwo}) the
Hamiltonian is
\begin{eqnarray}
H(\eta) &=& \int d^3\!k \sum_r \left\{
E_k(\eta) \left[ a_r^\dagger(k)a_r(k) - b_r(k)b_r^\dagger(k) \right]\right.\nonumber\\ 
&+&\left.
F_k(\eta)b_r(-k)a_r(k) 
+ F_k^*(\eta)a_r^\dagger(k)b_r^\dagger(-k)\right\} \ ,
\end{eqnarray}
where the equations of motion can be used to express $E_k$ and $F_k$
in terms of $u_+$ and $u_-$:\footnote{Here we choose the momentum $k$
along the third axis and use the representation in which $\gamma^3 =
\left( \begin{array}{cc} 0 & {\bf 1} \\ -{\bf 1} & 0 \end{array}
\right)$.}
\begin{eqnarray}
E_k & = & k {\rm Re}(u_+^*u_-) + 
am_{{\rm eff}}\left(1-\left|u_+\right|^2\right) \ , \nonumber \\
F_k & = & \frac{k}{2}(u_+^2-u_-^2) +am_{{\rm eff}} u_+u_- \ .
\end{eqnarray}

In order to calculate particle production one wants to write the
Hamiltonian in terms of creation and annihilation operators that are
diagonal. To do this one defines a new set of creation and
annihilation operators, $\hat{a}$ and $\hat{b}^\dagger$, related to
the original creation and annihilation operators $a$ and $b^\dagger$
through the (time-dependent) Bogolyubov coefficients $\alpha_k$ and
$\beta_k$,
\begin{eqnarray}
\label{sy}
\hat{a}(k) & = & \alpha_k(\eta) a(k) + \beta_k(\eta)b^\dagger(-k) \ ,
		\nonumber \\
\hat{b}^\dagger(k) 
       & = & -\beta^*_k(\eta) a(k) + \alpha^*_k(\eta)b^\dagger(-k) \ . 
\end{eqnarray}
The Bogolyubov coefficients will be chosen to diagonalize the
Hamiltonian.  Using the fact that the canonical commutation relations
imply $|\alpha_k|^2 + |\beta_k|^2=1$, the choice
\begin{equation}
\label{choice}
\frac{\alpha_k}{\beta_k} = \frac{E_k+\omega}{F^*_k} \ , \quad
\left| \beta_k \right|^2 = \frac{|F_k|^2}{2\omega(\omega+E_k)} \ ,
\end{equation}
results in a diagonal Hamiltonian,
\begin{equation}
H(\eta) = \int d^3\!k \sum_r \omega_k(\eta) \left[
\hat{a}_r^\dagger(k)\hat{a}_r(k) + \hat{b}^\dagger_r(k)\hat{b}_r(k) 
\right] \ . 
\end{equation}

We define the initial vacuum state $| 0 \rangle$ such that ${a} |
0 \rangle = {b} | 0 \rangle = 0$.
The initial conditions corresponding to the no-particle state are
\begin{equation}
u_\pm(0)=  \sqrt{\frac{\omega\mp m_{{\rm eff}}a}{\omega}} \ ; \qquad
\dot{u}_\pm(0) = iku_\mp(0)\mp ia m_{{\rm eff}} u_\pm(0) \ . 
\end{equation}

The (quasi) particle number
operator ${\cal N} = \hat{a}_r^\dagger(k)\hat{a}_r(k)$ such that the
particle number density $n$   is (including the two degrees of freedom from
the spin) 
\begin{equation}
\label{number}
n_{1/2}(\eta) = \langle 0 \left | {\cal N}/V \right | 0 \rangle =
\frac{1}{\pi^2a^3(\eta)} \int_0^\infty dk \, k^2 
\left| \beta_k \right|^2.
\end{equation}

Let us now consider a quadratic superpotential $W=M_\phi\Phi^2/2$.
The supergravity potential (\ref{scap}) is easily computed for such superpotential. In the limit $|\Phi|\ll \mpl$ it reduces to $V=M^2_\phi\phi^2/2$, but we have retained its complete supergravity form in the numerical analysis.
It is useful to write the equation of motions in terms of dimensionless variables. We introduce the dimensioneless time $\tilde{\tau}=M_\phi\tau$, as well as the dimensionless field $X=\phi/\phi_0$, so that the scalar field
is normalized by the condition $X_0=1$. We define $\phi_0$ as the value of the scalar field at the moment when the oscillations start.

By solving the Einstein Eqs. (\ref{ein1}) and (\ref{ein2}) and the equation of motion for the scalar field, we have  found  the time-dependent evolution of $m_{{\rm eff}}$. It is plotted in Fig. 2 in units of $M_\phi$ and for $\phi_0/\mpl=10^{-1}$. Notice in particular that at large times, $m_{{\rm eff}}$ tends to $M_\phi$. This is expected since one can  verify that,  in the limit of $|\Phi|\ll \mpl$, 
$m_{{\rm eff}}\simeq -\dot{\varphi}/a$ tends to  $\partial_\Phi^2 W=M_\phi$. 

\begin{figure}
\centerline{\leavevmode\epsfysize=8cm \epsfbox{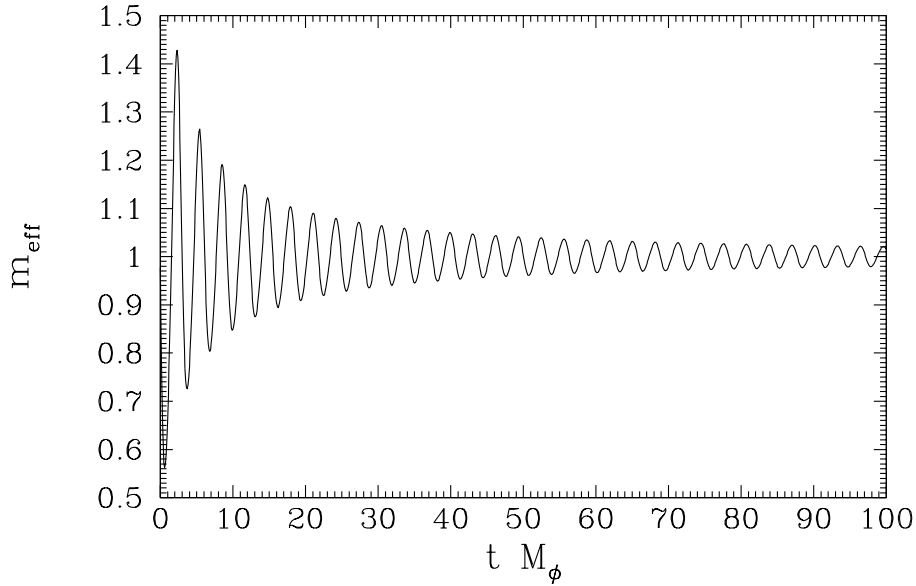}}
\vspace{10pt}
\caption{The evolution of the parameter $m_{{\rm eff}}$ in units of $M_\phi$ as a function of time and for a quadratic superpotential.  The initial condition is $\phi_0/\mpl=10^{-1}$.}
\label{meff_f=0.1}
\end{figure}


The result of our numerical integration for the power spectrum of helicity $\pm 1/2$ gravitinos is summarized in Fig. 3 for two different values of initial conditions.  Since $m_{{\rm eff}}$ changes by an amount $\sim M_\phi$ in a time scale $\sim M_\phi^{-1}$, one expects $k_{{\rm max}}\sim M_\phi$. This expectation is confirmed by our numerical results which indicates a cut-off in the spectrum for $k\sim M_\phi$.

\begin{figure}
\centerline{\leavevmode\epsfysize=8cm \epsfbox{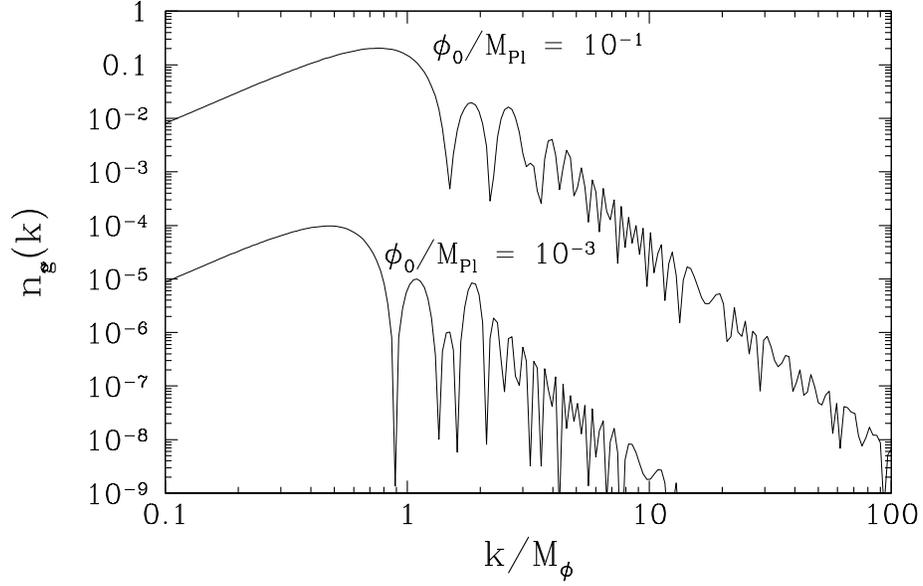}}
\vspace{10pt}
\caption{The power spectrum of helicity $\pm 1/2$ gravitinos for the initial conditions $\phi_0/\mpl=10^{-1}$ and $\phi_0/\mpl=10^{-3}$ and for a quadratic superpotential.}
\label{pn}
\end{figure}


Finally the ratio of the number density of gravitinos in units of the entropy density $s$ is given in Fig. 4 for $\phi_0/\mpl=10^{-1}$ and in units of the reheat temperature $T_{RH}$. If the  mass of the inflaton field is $M_\phi \simeq 10^{13}$ GeV as required by the normalization of density perturbations, we see that the non-thermal particle production of helicity $\pm 1/2$ gravitinos gives rise to a number density  well beneath the bound (\ref{lll})
\cite{linde,noi}.

\begin{figure}
\centerline{\leavevmode\epsfysize=8cm \epsfbox{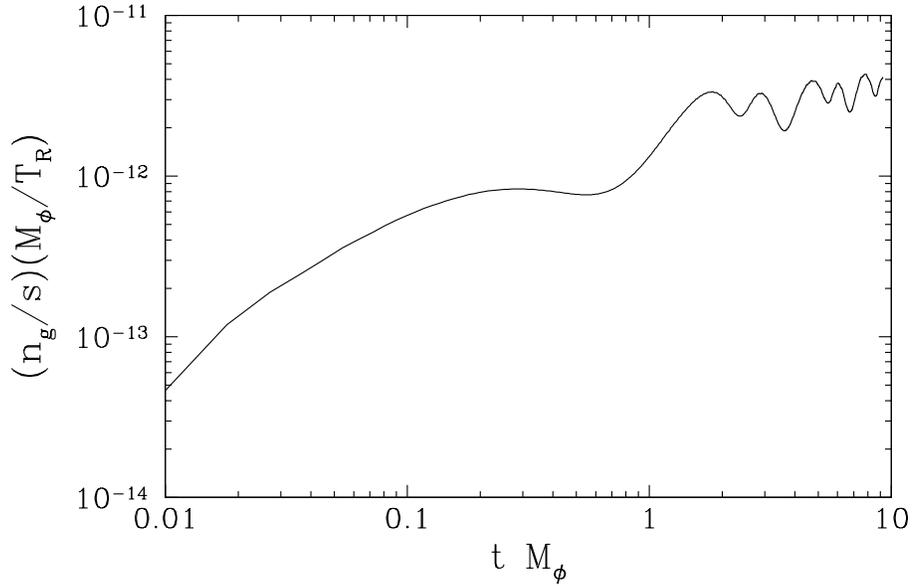}}
\vspace{10pt}
\caption{The ratio $n_g/s$ as a function of time for $\phi_0/\mpl=10^{-1}$ and for a quadratic superpotential.}
\label{nover}
\end{figure}


\begin{figure}
\centerline{\leavevmode\epsfysize=8cm \epsfbox{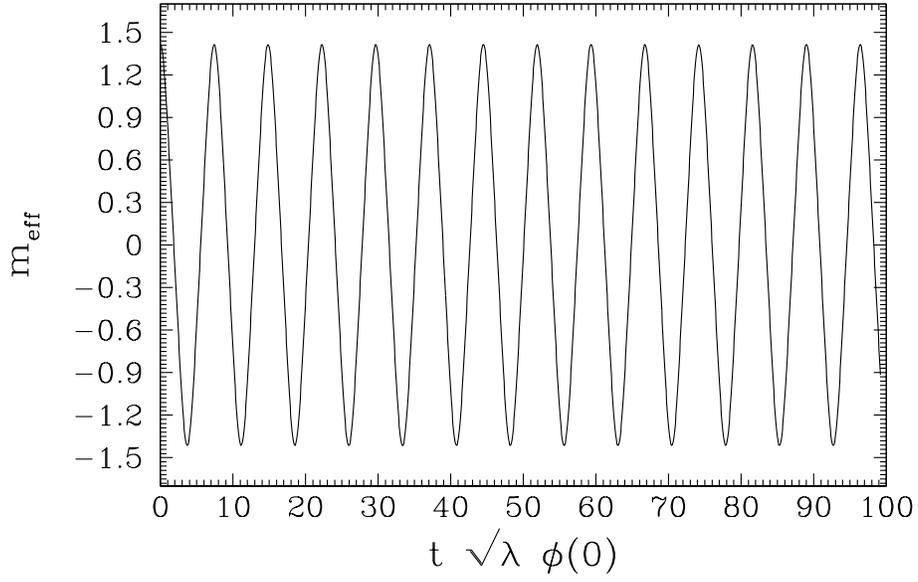}}
\vspace{10pt}
\caption{The evolution of the parameter $m_{{\rm eff}}$ as a function of time and for a cubic  superpotential.}
\label{lambdacase}
\end{figure}


\begin{figure}
\centerline{\leavevmode\epsfysize=8cm \epsfbox{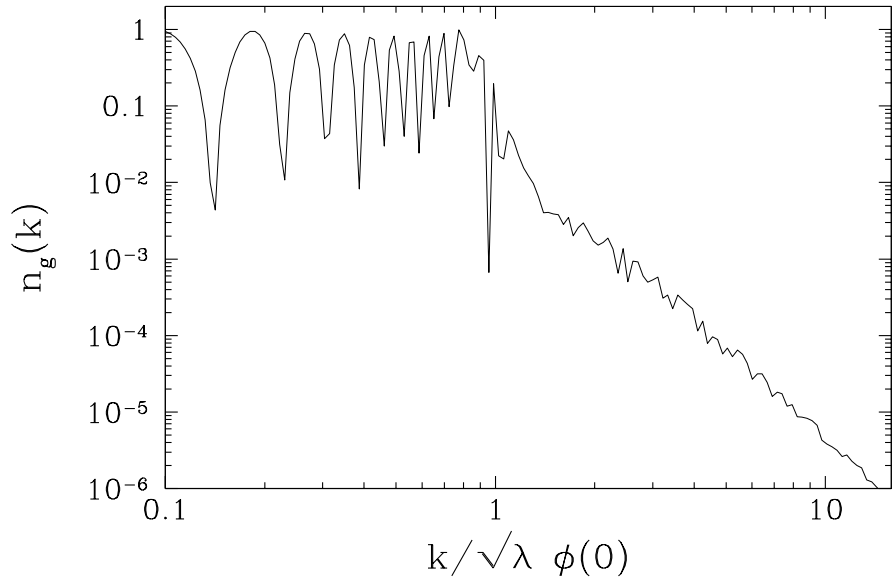}}
\vspace{10pt}
\caption{The power spectrum of helicity $\pm 1/2$ gravitinos for a cubic superpotential.}
\label{pnlambda}
\end{figure}


\begin{figure}
\centerline{\leavevmode\epsfysize=8cm \epsfbox{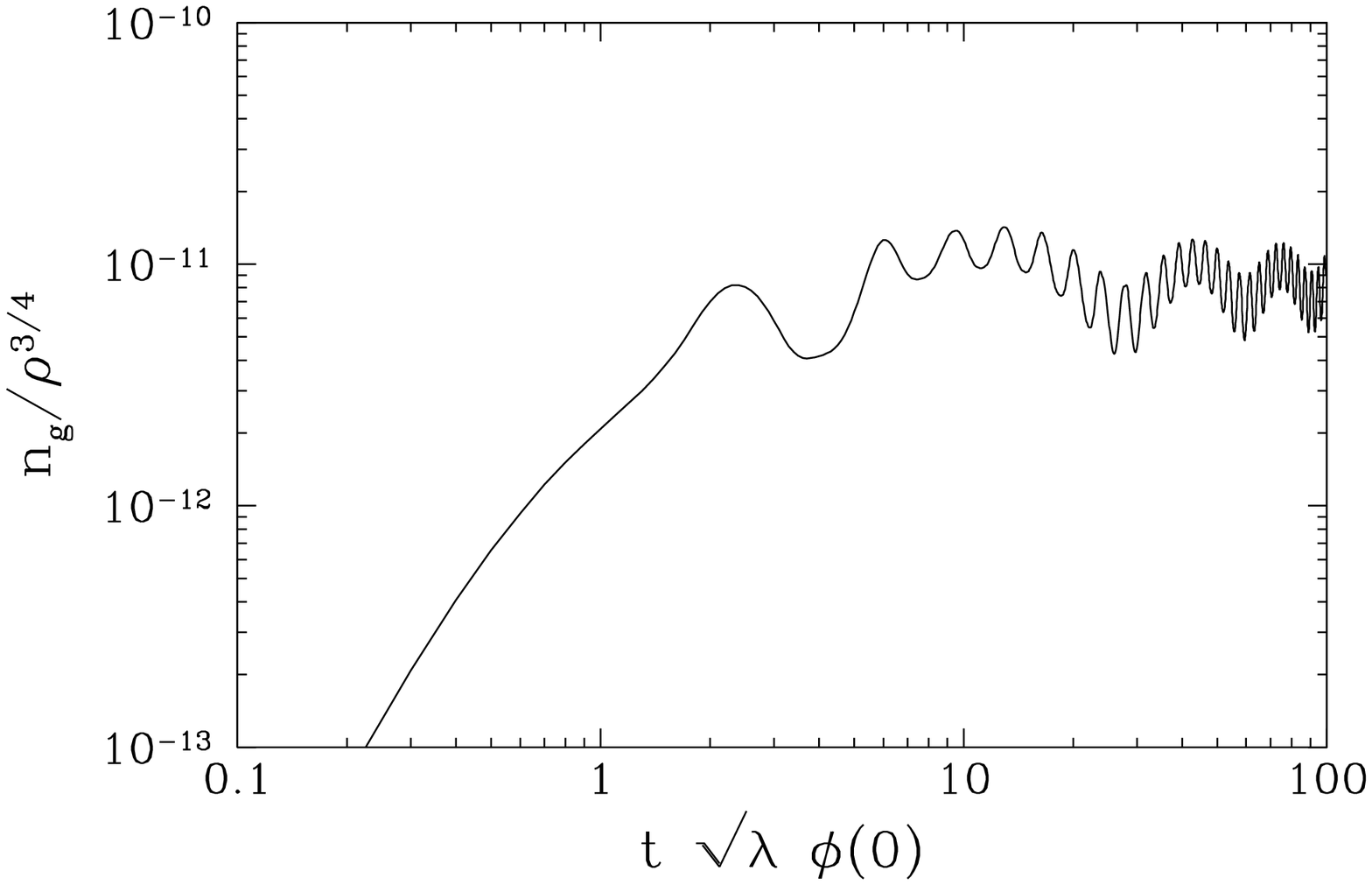}}
\vspace{10pt}
\caption{The ratio $n_g/\rho^{3/4}$ as a function of time for a cubic superpotential.}
\label{prova}
\end{figure}

Let us now consider a cubic superpotential $W=\sqrt{\lambda}\Phi^3/3$.
 In the limit $|\Phi|\ll \mpl$ the potential (\ref{scap}) reduces to $V=\lambda\phi^4/4$. A special feature of this theory is that the problem of gravitino production in an expanding universe can be completely reduced to a similar problem in Minkowski space-time by a simple conformal redefinition of the scalar field. This explains why the effective mass $m_{{\rm eff}}$ does not decrease with time, see Fig. 5. Furthermore, $m_{{\rm eff}}$ is expected to oscillate with maximum amplitude $\left|\partial_\Phi^2 W\right|=\sqrt{2}\sqrt{\lambda}\phi_0$ \cite{linde,noi}. This behaviour is well-confirmed by the numerical results given in Fig. 5.

The result of our numerical integration for the power spectrum of helicity $\pm 1/2$ gravitinos is summarized in Fig. 6. In this case $m_{{\rm eff}}$ changes by an amount $\sqrt{\lambda}\phi_0$ in a time scale $(\sqrt{\lambda}\phi_0)^{-1}$ and  one expects $k_{{\rm max}}\sim \sqrt{\lambda}\phi_0$. This expectation is again confirmed by our numerical results which indicated a cut-off in the spectrum for $k\sim \sqrt{\lambda}\phi_0$.

Finally, the ratio of the number density of gravitinos in units of the entropy density $\rho^{3/4}$ is given in Fig. 7.   Here $\rho$ indicate the energy density stored in  the massless oscillating scalar field $\phi$. The result $n_{3/2}/\rho^{3/4}$ will contradict the bound (\ref{lll}) by about one order of magnitude \cite{linde,noi} when the energy density in the scalar field is transferred to the energy density of a hot gas of relativistic particles.

\section{Non-thermal production of gravitinos in the  case of two chiral superfields}

As we already mentioned, constructing  a model of inflation in the context
 of  supergravity requires some effort. Realistic supersymmetric models of inflation require  the mass of the inflaton field
to be  much smaller than the Hubble rate. This is hard to achieve in the context of supergravity since supergravity corrections  spoil the flatness of the inflaton potential \cite{lr}. However, some exceptions are known and they usually involve more than one scalar field. 
  Consider the superpotential
\be
\label{sup}
W=  S\left(\kappa \frac{\phi^2}{2}- \mu^2\right),
\ee
where $\kappa$ is a dimensionless coupling of order unity \cite{sh,sugra}. The canonically-normalized inflaton field is
$\Phi\equiv\sqrt 2|S|$. The superpotential (\ref{sup}) leads to hybrid
inflation. Indeed, for  $\Phi\gg \Phi_c=\sqrt{2/\kappa}\mu $,
$\phi=0$ and the potential reduces to $V=\mu^4$ plus supergravity
and logarithmic corrections \cite{linderiotto}. If the K\"{a}hler potential for the superfield $S$ is minimal, the supergravity corections to the mass term
of the inflaton field cancel and they do not spoil the flatness of the potential. For $\Phi\gg \Phi_c$  the
Universe is trapped in the false vacuum and we have slow-roll inflation. The
scale $\mu$ is fixed to be around $5\times 10^{15}$ GeV to reproduce the observed
temperature anisotropy. 

When $\Phi=\Phi_c$, inflation ends because the false vacuum becomes unstable. The   
 field $\phi$  rapidly oscillates around  the minimum of
the potential at $\langle\phi\rangle=2\mu/\sqrt{\kappa}$, while the field
$\Phi$ rapidly oscillates around zero. The time-scale of the oscillations is
${\cal O}(\mu^{-1})$. The  mass scales at the end of inflation change by an
amount of order of $\mu$ within a time-scale $\sim \mu^{-1}$. Therefore, one
expects $k_{{\rm max}}\sim \mu$   and $n_{3/2}\sim 10^{-2}\mu^3$ \cite{noi}. After   reheating takes place, the final 
ratio $n_{3/2}$
to the entropy density is \cite{noi}
\be
\label{est}
\frac{n_{3/2}}{s}\sim 10^{-2}\frac{T_{RH}}{\mu}.
\ee
This violates the bound in Eq. (\ref{lll}) by at least four
orders of magnitude even if $T_{RH}\sim 10^9$ GeV and imposes a stringent
upper bound on the reheating temperature
$T_{RH}\simlt 10^5$ GeV \cite{noi}.

The estimate (\ref{est}) obtained in ref.  \cite{noi} was based on the assumption that the results on the  gravitino
production for a  single one chiral superfield model are valid 
in a  theory with  more than one superfield. In the following, we wish to 
show that this assumption is justified. Instead of attacking the problem of the production of helicity $\pm 1/2$ gravitinos in theories with more than one chiral superfield from a supergravity point of view, we make use of the 
gravitino-Goldstino equivalence theorem. The identification of the helicity $\pm 1/2$ gravitino with the Goldstino is well justified, since the amplitudes of the oscillating fields in the models of supersymmetric hybrid inflation are
far below the Planck scale. 

We generically denote the two chiral superfields involved in the generic problem at hand by $\Phi_1$ and $\Phi_2$ and by $\chi_1$ and $\chi_2$ the corresponding fermionic degrees of freedom. The combination $\hat{\chi}$ can be expressed, making use of Eq. (\ref{linear}), as
\beq
\hat{\chi}=\Theta_1^\dagger \chi_1^\perp+\Theta_2^\dagger \chi_2^\perp=-2i\gamma^0\frac{
\Delta}{\Theta_2}\chi_1^\perp,
\eeq
where we have defined $\Delta$ as
\beq
\Delta=W_1\dot{z}_2-W_2\dot{z}_1=-\frac{i}{2}\gamma^0\left(
\Theta_1 \Theta_2^\dagger-\Theta_2 \Theta_1^\dagger\right).
\eeq
Using Eq. (\ref{a2}), the master Eq. (\ref{master}) becomes
\beq
\label{master2}
\ddot{\eta}+k^2\eta + \alpha\dot{\eta}+i\gamma^0\hat{k}\left(\dot{{\cal G}}^\dagger+\alpha^\dagger {\cal G}^\dagger\right)\eta=0,
\eeq
where 
\beq
\alpha=i\frac{\gamma^0}{\Delta}\left(\Theta_2^\dagger \dot{\Theta}_1-
\Theta_1^\dagger \dot{\Theta}_2\right).
\eeq
Notice that $\Delta$ satisfy the following differential equation
\beq
\frac{\dot{\Delta}}{\Delta}=-\frac{\left(\alpha+\alpha^\dagger\right)}{2},
\eeq
which is solved by
\beq
\Delta=\Delta(0)e^{-\int_0^t \frac{\left(\alpha+\alpha^\dagger\right)}{2}}.
\eeq
Therefore during the time evolution of the system, $\Delta$ will never   vanish.

Redefining $\eta\rightarrow {\rm exp}\left(-\int dt\:\alpha/2\right)\eta$, Eq. (\ref{master2}) can be recast in the form
\beq
\label{master3}
\ddot{\eta}+k^2\eta - \left(\frac{\dot{\alpha}}{2}+\frac{\alpha^2}{4}\right)\eta+
i\gamma^0\hat{k}\left({\cal G}^\dagger e^{\int \alpha^\dagger dt}\right)^{\cdot}\frac{\Delta}{\Delta(0)}\eta=0.
\eeq
Finding  an exact solution to 
Eq. (\ref{master3}),  or even  studying  the problem numerically,    goes beyond the scope of this paper; we will  limit ourselves to 
outline  a standard  approximation method to estimate the number density of helicity $\pm 1/2$ gravitinos. If we define
$\eta_k^{{\rm in}}(t)\propto e^{-i\Omega t}$ to be  the solution of the equation at $t\rightarrow -\infty$, {\it i.e.} a plane-wave, 
Eq. (\ref{master3}) can be written as an integral equation
\beq
\label{master4}
\eta_k(t)=\eta_k^{{\rm in}}(t)+\Omega^{-1}\int_{-\infty}^t\: V_k(t^\prime)\:{\rm sin}\left[\Omega(t-t^\prime)\right]\eta_k(t^\prime)dt^\prime,
\eeq
where

\beq
V_k(t)=\left(\frac{\dot{\alpha}}{2}+\frac{\alpha^2}{4}\right)-
i\gamma^0\hat{k}\left({\cal G}^\dagger e^{\int \alpha^\dagger dt}\right)^{\cdot}\frac{\Delta}{\Delta(0)}.
\eeq
Decomposing $\eta_k(t)$ in terms of  $u_r^T\equiv[u_+(t)\psi_r(k), r u_-(t)\psi_r(k)]$ and $v_r^T \equiv [-r u^*_-(t)\psi_r(k),$  $u_+^*(t)\psi_r(k)]$, where $r$ is the spin index, 
in the late time region, Eq. (\ref{master4}) possesses the solution
\beq
u_r^{{\rm out}}(t)=\alpha^{rr'}_k u_{r'}^{{\rm in}}(t)+ \beta^{rr'}_k v^{{\rm in}}_{r'}(t),
\eeq
where the  Bogolyubov coefficient $\beta^{rr'}_k$ is given by
\beq
\beta^{rr'}_k=-(i/2\Omega)\int_{-\infty}^{\infty}\: \:e^{-2i\Omega t}v^{{\rm in}\dagger}_{r'}(t)
V_k(t) u_r(t)\:dt,
\eeq
and we  have let $t\rightarrow+\infty$.
If we treat $V_k(t)$ as a perturbative potential, then we can solve Eq. (\ref{master4}) by iteration. To the lowest order in $V_k$, one has 
$u_r(t)=u_r^{{\rm in}}(t)$ and the Bogolyubov coefficient $\beta^{rr'}_k$
becomes

\beq
\beta^{rr'}_k=-(i/2\Omega)\int_{-\infty}^{\infty}\:e^{-2i\Omega t}
v^{{\rm in}\dagger}_{r'}(t)
V_k(t) u^{{\rm in}}_r(t)\:dt.
\eeq
The corresponding number of Goldstinos (or, equivalently,  helicity $\pm 1/2$ gravitinos) in a given spin state is therefore
\beq
N_{kr}=\sum_{r'}|\beta_k^{rr'}|^2.
\eeq
Even though this approximation  is expected to offer only part
of the information about   the resonant behaviour of the system,  we believe it
provides  the right order of magnitude for the number density. In typical supersymmetric hybrid models of inflation, like the one described by the superpotential (\ref{sup}),  the
system relaxes to the minimum in a time-scale   much shorter than the Hubble
time $\sim H_I^{-1}$, since the  frequency is 
set  by the height of the potential
$V^{1/4}\sim \mu \gg H_I$ during inflation. The  number of particles $N_{kr}$
depends upon $\widetilde V_k(\Omega)$, the Fourier transform of $V_k(t)$. 
Since   $V_k(t)$ changes by an amount $\sim \mu$ in a timescale $\sim\mu^{-1}$,  
$\widetilde V_k(\Omega)$  rapidly dies out for frequencies $\Omega\gg \mu$, $
\widetilde V_k(\Omega)\propto 1$ for $\omega\simlt \Omega_{{\rm max}}\sim \mu$ and zero otherwise. The 
 number of helicity $\pm 1/2$  will be  $\sim\Omega_{{\rm max}}^3\sim \mu^3$, confirming the original  estimate  made  in ref. \cite{noi}.

\vskip1cm
\centerline{\large\bf Acknowledgements}
\vskip 0.2cm

We would like to thank F. Feruglio, R. Kolb and A. Linde for discussions.


\vskip1cm
\begin{thebibliography}{99}
%

%
\bibitem{sugra} For a review, see H.P. Nilles, Phys. Rep.
{\bf 110}, 1  (1984).

\bibitem{nucleo} D. Lindley, Ap. J. {\bf 294}, 1  (1985);\\ J. Ellis {\it
et al.}, Nucl. Phys. {\bf 259}, 175 (1985);\\ S. Dimopoulos {\it et al.}, Nucl.
Phys. {\bf B311}, 699 (1988);\\ J. Ellis {\it et al.}, Nucl. Phys. 
{\bf B373}, 399 (1992). 

\bibitem{ellis} J. Ellis, A. Linde, and D. Nanopoulos, Phys. Lett. {\bf
B118}, 59 (1982); \\ D. Nanopoulos, K. Olive, and M. Srednicki, 
Phys. Lett. {\bf
B127}, 30 (1983);\\ J. Ellis, J. Kim, and D. Nanopoulos, 
Phys. Lett. {\bf B145}, 181 
(1984).


\bibitem{kaw} M. Kawasaki and T. Moroi, 
Prog. Theor. Phys. {\bf 93}, 879 (1995).
 

\bibitem{linde} R. Kallosh, L. Kofman, A. Linde and  A. Van Proeyen, hep-th/9907124. 


\bibitem{noi} G.F. Giudice, I. Tkachev and  A. Riotto, JHEP 9908:009 (1999).

\bibitem{porc}  D.H. Lyth, D. Roberts, and M. Smith,  
Phys. Rev. {\bf D57}, 7120 (1998);  \\ 
A.L. Maroto and A. Mazumdar, hep-ph/9904206; \\
M.~Lemoine,
Phys.\ Rev.\ {\bf D60}, 103522 (1999).
 

\bibitem{new} D.H. Lyth, hep-ph/9911257.

\bibitem{eq} P. Fayet, Phys. Lett. {\bf B175}, 471 (1986);\\ R.~Casalbuoni, S.~De Curtis, D.~Dominici, F.~Feruglio and R.~Gatto,
Phys.\ Lett.\ {\bf B215}, 313 (1988); \\
R.~Casalbuoni, S.~De Curtis, D.~Dominici, F.~Feruglio and R.~Gatto,
Phys.\ Rev.\ {\bf D39} (1989) 2281.





















%
\bibitem{lr} See, for example, D.H. Lyth and A. Riotto, 
{\it Models of inflation, particle physics and
the spectral index of the density perturbations},     Phys. 
Rept. {\bf 314} (1999) 1.


\bibitem{book} 
E. W. Kolb and M. S. Turner, 
        {\it The Early Universe}, (Addison-Wesley, Menlo Park, Ca., 1990). 

\bibitem{kls} L.~Kofman, A.~Linde and A.A.~Starobinsky,
Phys.\ Rev.\ Lett.\ {\bf 73}, 3195 (1994).


\bibitem{kt1} S.Y.~Khlebnikov and I.I.~Tkachev,
Phys.\ Rev.\ Lett.\ {\bf 77}, 219 (1996).



\bibitem{kt2} S.Y.~Khlebnikov and I.I.~Tkachev,
Phys.\ Rev.\ Lett.\ {\bf 79}, 1607 (1997).





\bibitem{review} For a recent review, see A. Riotto and A. Trodden, {\it Recent progress in baryogenesis}, hep-ph/9901362, to appear in Anual Review of Nuclear and Particle Science; A. Riotto, {\it Theories of Baryogenesis}, Lectures given  at the  ICTP Summer School in High-Energy Physics and Cosmology, Miramare, Trieste, Italy, 29 Jun - 17 Jul 1998, hep-ph/9807454. 



\bibitem{turner} 
R. J. Scherrer and M. S. Turner, 
        Phys. Rev. {\bf D31}, 681 (1985).


\bibitem{ckr} D.J.H. Chung, E.W. Kolb and  A. Riotto,   Phys. Rev. {\bf D60} (1999) 063504.



\bibitem{vol} D.V. Volkov and V.A. Soroka, JETP {\bf 18} (1973) 312;\\ S. Deser
and B. Zumino, Phys. Rev. Lett. {\bf 38} (1977) 312.

\bibitem{cre} E. Cremmer et {\it al}, Nucl. Phys. {\bf 147} (1979) 105.


\bibitem{ferm}J. Baacke, K. Heitmann, and C. Patzold, Phys. Rev. {\bf D58},
125013 (1998); \\ P. B. Greene and L. Kofman, Phys. Lett. {\bf B448}, 6 
(1999);\\
 V.A. Kuzmin and I.I. Tkachev,
Phys. Rev. {\bf D59}, 123006 (1999);\\
G.F. Giudice,  M. Peloso, A. Riotto, and I. Tkachev, JHEP 9908:014 (1999):\\
D.J.~Chung, E.W.~Kolb, A.~Riotto and I.I.~Tkachev,
hep-ph/9910437.


 



\bibitem{linderiotto}  
A. D. Linde and A. Riotto,  
Phys. Rev. {\bf D56}, 1841 (1997). 


\bibitem{dterm} P. Binetruy and G. Dvali, Phys. Lett. {\bf B388}, 241 (1996);\\
E. Halyo, Phys. Lett. {\bf B387}, 43 (1996); \\ D.H. Lyth and A. Riotto,  Phys.
Lett. {\bf 412}, 28 (1997); \\ G. Dvali and A. Riotto,  Phys.
Lett. {\bf 417}, 20 (1998); \\ J.R. Espinosa, A. Riotto, and G.G.
Ross, Nucl. Phys. {\bf B531}, 461 (1998);\\ S.F. King 
and  A. Riotto, Phys. Lett. {\bf B442}, 68 (1998). 


\bibitem{gel} J.A. Casas, G.B. Gelmini, and A. Riotto, hep-ph/9903492.

\bibitem{sh} G. Dvali,  Q. Shafi, and R. Schaefer,  
Phys. Rev. Lett. {\bf 73}, 1886 (1994). 

\bibitem{sugra} A.~Linde and A.~Riotto,
Phys.\ Rev.\ {\bf D56}, 1841 (1997).


\end{thebibliography}
\end{document}